\magnification=\magstep1
\raggedbottom
\hsize=165 true mm
\vsize=205 true mm

\font\eightrm=cmr8
\font\eighti=cmmi8
\font\eightsy=cmsy8
\font\eightit=cmti8
\font\sixi=cmmi6
\font\sixsy=cmsy6
\font\eightbf=cmbx8

\def\singlespace{\baselineskip=12pt}  
\def\sesquispace{\baselineskip=14pt}

\def\tilde{\widetilde}
\def\M{{\cal M}}
\def\N{{\cal N}}
\def\O{{\cal O}}
\def\U{{\cal U}}
\def\P{{\cal P}}
\def\Q{{\cal Q}}
\def\Reals{{\rm I\!\rm R}}
\def\cent{\centerline}
\def\ra{\rangle}
\def\la{\langle}
\def\lla{\langle\langle}
\def\rra{\rangle\rangle}
\def\dal{\hbox{$\/\vrule height1.55ex width0.15ex depth.1ex \hskip-0.15ex
                  \vrule height0.10ex width1.4ex depth.1ex
                  \vrule height1.55ex width0.15ex depth.1ex \hskip-1.5ex
                  \vrule height1.55ex width1.4ex depth-1.4ex \,$}}

\def\implies{\Rightarrow}
\def\ideq{\equiv}
\def\bar{\overline}

\def\S{{\cal S}}
\def\X{{\cal X}}
\def\O{{\cal O}}
\def\B{{\cal B}}
\def\K{{\cal K}}
\def\Y{{\cal Y}}

\sesquispace

\cent {\bf A Causal Order for Spacetimes with $C^0$ Lorentzian Metrics:}
\cent {\bf Proof of Compactness of the Space of Causal Curves}\footnote{}
{Published {\it Class. Quant. Grav.}{\bf 13}: 1971-1994 (1996); 
 e-print archive: gr-qc/9508018.}
\cent{\phantom {by}}
\cent{R.D. Sorkin}
\cent{Dept. of Physics, Syracuse University, Syracuse, NY13244--1130, USA}
\cent{and}
\cent{Instituto de Ciencias Nucleares, UNAM,
      A. Postal 70-543
      M\'exico, D.F. 04510, M\'exico}
\cent{E-Mail: rdsorkin@mailbox.syr.edu}
\smallskip
\cent{\it and}
\smallskip
\cent{E. Woolgar}
\cent{Dept. of Mathematics, University of Saskatchewan, 106 Wiggins Ave.,}
\cent{Saskatoon, SK, Canada S7N 5E6}
\cent{and}
\cent{Winnipeg Institute for Theoretical Physics, Dept. of Physics, University
of Winnipeg,}
\cent{515 Portage Ave., Winnipeg, MB, Canada R3B 2E9}
\cent{E-Mail: woolgar@math.usask.ca }

\bigskip\leftskip=1.5truecm\rightskip=1.5truecm
\singlespace
\cent{\bf Abstract}
\smallskip
\noindent
We recast the tools of ``global causal analysis'' in accord with an
approach to the subject animated by two distinctive features: a
thoroughgoing reliance on order-theoretic concepts, and a utilization of
the Vietoris topology for the space of closed subsets of a compact set.  We
are led to work with a new causal relation which we call $K^+$, and in
terms of it we formulate extended definitions of concepts like causal
curve and global hyperbolicity.  In particular we prove that, in a
spacetime $\M$ which is free of causal cycles, one may define a causal
curve simply as a compact connected subset of $\M$ which is linearly
ordered by $K^+$.  Our definitions all make sense for arbitrary $C^0$
metrics (and even for certain metrics which fail to be invertible in
places).  Using this feature, we prove for a general $C^0$ metric, the
familiar theorem that the space of causal curves between any two compact
subsets of a globally hyperbolic spacetime is compact.  We feel that our
approach, in addition to yielding a more general theorem, simplifies and
clarifies the reasoning involved.  Our results have application
in a recent positive energy theorem [1], and may also prove useful in
the study of topology change [2].  We have tried to make our
treatment self-contained by including proofs of all the facts we use which
are not widely available in reference works on topology and differential
geometry. 
\bigskip\leftskip=0truecm\rightskip=0truecm

\sesquispace

\vskip 2 true cm
\noindent
\cent{\bf 1. Introduction}
\par\nobreak                
\noindent
Global geometrical properties of manifolds often condition the behavior
of curves within the manifold.  For example, in a riemannian manifold,
the metric property of completeness implies the existence of a minimum
length (and therefore geodesic) curve joining any two points of the
manifold (Hopf-Rinow theorem [3]).  In a lorentzian manifold, the
situation is somewhat different.  Here properties of order and topology
tend to take precedence over strictly metric properties, and one is
primarily interested, not in arbitrary curves, but in causal ones.  In
the lorentzian case there are two analogues of a minimal geodesic: a
longest timelike curve and ({\it cf.} [4], [1]) a ``fastest causal
curve'' (necessarily a null geodesic).  An analogue of the Hopf-Rinow
result does exist, but the causal property of global hyperbolicity
replaces that of Cauchy completeness as the basic assumption.  Moreover
the compactness of the space of causal curves between two points of the
manifold now becomes an essential part of the argument.  This
compactness thus plays a crucial role in the global analysis of
lorentzian manifolds.  In particular, many of the singularity theorems
of general relativity rely directly on it, as does a recent proof of
positivity of the total energy in general relativity [1].

For this last application however, the standard compactness theorem does
not suffice, because it presupposes the existence of convex normal
neighborhoods, which in turn requires that the metric be of
differentiability class $C^2$ (or at least $C^{2-}$).\footnote{*}
{whereas the proof in [1] takes place in conformally extended
spacetime, which is not smooth at the point at spatial infinity, $i^0$.}
On the other hand, since the compactness theorem itself envisions causal
curves which are not even differentiable, one might expect that this degree
of smoothness should not be necessary.  Below we will confirm this
expectation by proving a compactness theorem which holds even when the
metric is only $C^0$.

In fact, most of our results will be meaningful and true even when the
metric is degenerate in places, and thus is not strictly lorentzian (nor
even strictly $C^0$ insofar as the inverse metric is concerned).  This may
be of importance in the study of topology change, because topology-changing
manifolds (non-product cobordisms) in general require such degeneracies
[2].

In making do without recourse to smoothness assumptions, we are led to a
set of definitions and proofs which rely more directly on order-theoretic
ideas than has been customary heretofore.  The resulting framework is not
only of wider applicability than previous ones, but we believe it affords a
proof of our main theorem which is conceptually simpler as well.  
In the same way, we suspect that the bulk of the theory of ``global causal
analysis'' could instructively be recast along such order-theoretic and
general-topological lines, the main attraction in doing so being the
prospect of a more elementary and more perspicuous development.   This, and
not greater generality {\it per se}, is what primarily motivated the
work reported herein.  (Indeed, if accommodation to $C^0$ metrics were the
only goal, we suspect that existing approaches could be adapted to that end
without undue fuss.)

Two technical modifications animating our treatment are the use of an
unfamiliar topology in studying convergence of families of curves, and the
introduction of a new causal relation (which we have dubbed $K^+$ in
analogy with the notations $I^+$ and $J^+$) in terms of which our notions
of causal curve and of global hyperbolicity are defined.  The main
advantage of $K^+$ in comparison with $I^+$ and $J^+$ is that it is
topologically closed as well as transitive, which helps 
for example in the proofs of Lemmas 14 and 15 and Theorems 20 and 21
below.  The topology just alluded to is the so-called Vietoris topology,
which is naturally defined not just on curves, but on the space $2^{\X}$ of
all closed non-empty subsets of a topological space ${\X}$.  Its primary
advantage is the availability of the compactness result, Theorem 1, but it
has as well the minor virtue that the space of causal curves becomes
automatically Hausdorff in a natural way, even when the endpoints are not
held fixed.  We note that the use of the Vietoris topology in proving
compactness might be adapted to spaces of more general objects than causal
curves (null hypersurfaces, closed future/past sets, {\it etc.}), although
we will not pursue this here.

An outline of the remainder of this paper is as follows.  Section 2 is a
summary of the general facts we will need involving ordered sets on one
hand, and the Vietoris topology on the other hand.  The key properties of
the latter are that a Vietoris limit of connected subsets of a bicompact
space is connected, and that $2^\M$ is bicompact whenever $\M$ itself is
(bicompact $\ideq$ compact and Hausdorff).  In connection with ordered
spaces, we give general criteria for a poset to be isomorphic to the real
unit interval $[0,1]$, and for a space which carries both an order and a
topology to be simultaneously isomorphic to $[0,1]$ in both respects.
Complete proofs for these results will be found in Appendix A.  

Section 3 is in a sense the heart of the paper.  In it we define our basic
causality relation $K^+ = \; \prec$ to be the smallest closed\footnote{*}
{Closure here means topological closure with respect to the manifold
topology.  One could also entertain an order-theoretic notion of closure by
letting $x_n \nearrow x$ (respectively $x_n \searrow x$) mean that $x$ is
the supremum (respectively infimum) of the $x_n$, and defining closure in
terms of this kind of convergence.  This possibility raises the prospect 
that order will eventually absorb both geometry {\it and} topology in
the basic formulations of the subject.}
and transitive relation containing $I^+$; and based on this definition, we
introduce the elements of our framework for ``$C^0$ causal analysis'',
including: a causality condition called ``$K$-causality'' which requires
that $K^+$ be a partial order; a generalized definition of causal curve and
of global hyperbolicity; the fact that a $K$-causal manifold is
automatically ``locally causally convex''; and a simple characterization of
a causal curve --- as a compact connected subset linearly ordered by $K^+$
--- which is valid in an arbitrary $K$-causal spacetime, and which we use
when proving our main theorem in the following section.  The proof of the main
theorem itself in Section 4 is short and straightforward, and amounts to
showing that the Vietoris limit of a sequence of causal curves is itself a
causal curve.  This being demonstrated, the compactness properties of the
Vietoris topology lead directly to the desired conclusion, namely that the
space of causal curves between two compact subsets of a globally hyperbolic
$C^0$-lorentzian manifold is itself compact.  

Appendix B is devoted to a comparison of our definitions with ones which
are standard in the $C^2$ situation.  In particular we prove that, when
restricted to $C^2$ metrics, our definitions of causal curve and of global
hyperbolicity of a manifold agree with the usual ones.

Concerning notation, a $C^p$-{\it lorentzian manifold} $\M$ means herein a
time-oriented (paracompact) manifold of differentiability class $C^{p+1}$
($p \ge 0$), equipped with a symmetric tensor field $g_{ab}$ of
differentiability class $C^p$ and signature $(-1,1,1,1,\dots)$.  (A
time-orientation for such a manifold can be given by choosing a
nowhere-vanishing timelike vector field $u^a$ --- provided one exists.)
Any manifold which occurs in this paper will be assumed, without explicit
mention, to be at least $C^0$-lorentzian.  We use the abstract index
convention, denoting such indices by early Latin letters.
Finally, we use throughout the convention that the symbol $\Gamma$ denotes 
the image of $\gamma$, where $\gamma : [0,1]\to\M$ is a {\it path} or
parameterized curve.

Finally, we note here that none of our considerations will actually depend
on making a choice of representative metric within a given conformal
equivalence class.

\vskip 2 true cm
\noindent
\cent{\bf 2. Some Definitions and Theorems Concerning}
\cent {\bf   Ordered Sets and the Vietoris Topology }
\par\nobreak\smallskip  
\noindent
In this preliminary section we assemble some definitions and theorems of a
general character which will be useful later.  As none of these theorems
relate specifically to lorentzian manifolds, we do not prove them here;
but, for completeness, and as some of the theorems are not easily 
located in the literature, we do give full proofs in Appendix A.

We begin with the Vietoris topology, which is a topology on the space
$2^{\cal X}$ of non-empty closed subsets of an arbitrary topological space
${\cal X}$.  Later we will characterize it in terms of convergence; here we
define it by giving the open sets directly, or rather by giving
a {\it base} for them (a base being a collection of open sets whose
arbitrary unions furnish the topology itself).  
Let $\{ {\cal A}_0,{\cal A}_1,\dots , {\cal A}_n\} $ be a finite collection
of open subsets of ${\cal X}$.  We define 

{\narrower\smallskip\noindent
  ${\cal B}({\cal A}_0;{\cal A}_1,{\cal A}_2,\dots ,{\cal A}_n)
  = 2^{{\cal A}_0}\cap
 \big(2^{\cal X}\backslash 2^{{\cal X}\backslash{\cal A}_1} \big )
 \cap \big ( 2^{\cal X} \backslash 2^{{\cal X}\backslash {\cal A}_2} \big )
 \cap \dots \cap
 \big( 2^{\cal X} \backslash 2^{{\cal X}\backslash {\cal A}_n} \big) \quad ,$
\smallskip}

\noindent
where ${\cal A}_1\backslash {\cal A}_2$ is the set of all elements of
${\cal A}_1$ which are not elements of ${\cal A}_2$.  Equivalently, if
${\cal C}$ is a closed subset of ${\cal X}$ then

{\narrower\smallskip\noindent
${\cal C}\in {\cal B}({\cal A}_0;{\cal A}_1,{\cal A}_2,\dots ,{\cal A}_n)
 \Longleftrightarrow 
{\cal C} \subseteq {\cal A}_0$ and ${\cal C}$ meets ${\cal A}_i$ 
for $i=1,\dots,n\quad .$
\smallskip}

\noindent
In other words, ${\cal C}$ must be large enough to intersect each of the
${\cal A}_i$ and small enough never to stray outside of ${\cal A}_0$.  The
sets ${\cal B}({\cal A}_0;{\cal A}_1,{\cal A}_2,\dots , {\cal A}_n)$ for
finite $n$ constitute a base for the Vietoris topology on $2^{\cal X}$.  We
can also give a somewhat simpler {\it sub-base} for this topology. It
consists simply of sets of the form $B({\cal A};{\cal X})$ {\it together
with} sets of the form $B({\cal X};{\cal A})$.  The former are the sets
whose elements are the closed subsets ${\cal C}\subseteq {\cal A}$, ${\cal
A}$ being an open subset of ${\cal X}$. The latter are the sets of closed
sets ${\cal C} \subseteq {\cal X}$ that meet the open set ${\cal
A}\subseteq {\cal X}$.  The finite intersections of sets of these two types
produce the base sets ${\cal B}({\cal A}_0;{\cal A}_1, {\cal
A}_2,\dots,{\cal A}_n)$.  Note that the empty subset of ${\cal X}$ is not
an element of the space $2^{\cal X}$, \footnote{$^{\ddag}$}
{ Because of this, the notation $2^{\cal X}-1$ might be more evocative,
  albeit more clumsy.  Still another notation is ${\cal F}_0({\cal X})$,
  which is used in [5]. }
and correspondingly, the limit of a sequence of non-empty closed sets is
always non-empty itself.  We will refer to limits with respect to the
Vietoris topology as {\it Vietoris limits}.

The next two propositions, both intuitively plausible, will feed
directly into the proof of the  compactness theorem which is our
principal result in Section 4.

{\narrower\smallskip\noindent
{\bf Theorem 1:}
  The topological space $2^{\cal X}$ 
  is bicompact whenever ${\cal X}$ itself is. 
\smallskip}

\vbox{
{\narrower\smallskip\noindent
{\bf Lemma 2:}

\item{(i)}{In a compact space, every Vietoris limit is a compact set.}
\item{(ii)}{In a bicompact space, a Vietoris limit of connected sets is
connected.}

\smallskip}}

In addition to the use of the Vietoris topology, our approach rests on a
liberal employment of order-theoretic concepts.  For that reason, we gather
here a number of definitions relating to ordered sets, and we state two
general theorems giving criteria for an ordered space to be isomorphic to
the closed unit interval $[0,1]$.  Both these theorems are proved in 
Appendix A. 

{\narrower\smallskip\noindent
{\bf Definition 3:}
 A {\it partial order} is a relation $\prec$ defined on a set $S$ which
   satisfies the axioms of
 
\item{(i)}{asymmetry: $p\prec q$ and $q\prec p$ $\Rightarrow$ $p=q$, and}
\item{(ii)}{transitivity: $p\prec q$ and $q\prec r$ $\Rightarrow$ $p\prec r$.}
 
\noindent
The partial order is {\it reflexive} iff it
 observes the convention that every element
 of $S$ precedes itself ($p\prec p\ \forall p\in S$), and
 {\it irreflexive} iff it observes the
 convention that no element precedes itself.  (In the
 irreflexive case, condition (i) would be more naturally stated just as the
 condition that  $p\prec q$ and $q\prec p$ never occur together.)
\smallskip}

\noindent
A partial order is often called simply an {\it order}, and a set endowed
with an order is called an {\it (partially) ordered set} or {\it poset}.

{\narrower\smallskip\noindent
{\bf Definitions 4:}
 Let $P$ be a reflexive poset with the order relation denoted by $\prec$.  A
{\it linearly ordered} subset or {\it chain} $Q$ in $P$ is a subset such
that $x,y\in Q\Rightarrow$ either $x\prec y$ or $y\prec x$ (as is the case
with $Q=\Reals$, for example).  The corresponding order is called a {\it
linear order} (also a {\it total order}).  For any $x, y \in P$, we define
the {\it order-open interval} $\la x,y \ra$ to be 
$\{ r| x\prec r\prec y,\ |x\neq r\neq y\}$ and 
the {\it order-closed interval} $\lla x,y \rra$ to be 
simply $\{ r| x\prec r\prec y\}$, with the half-open
intervals $\lla p,q \ra$ and $\la p,q \rra$ defined analogously.\footnote{*}
{ The terms ``open'' and ``closed'' seem appropriate for the linear
  orders we will be dealing with in this paper.  More generally, the words
 ``exclusive'' and ``inclusive'' might be more descriptive. }
A {\it minimum} element of a poset (necessarily unique)
is one which precedes every other
element, and dually for a {\it maximum} element.  We will always denote the
former by $0$ and the latter by $1$.  For subsets $A$ and $B$ of $P$, we
write $A\prec B$ to mean that $a\prec b \; \forall a\in A, b\in B$.  A
subset $S\subseteq P$ is {\it order-dense} in $P$ iff
it meets every non-empty order-open interval $\la x,y \ra$.  
A function $f:X\to Y$, for posets $X$ and $Y$, is a {\it tonomorphism} or
{\it order-isomorphism}, and we say that $X$ and $Y$ are {\it tonomorphic}
or {\it order-isomorphic} via $f$, iff $f$ is an order-preserving
bijection, in which case we have, clearly,
$x_1 \prec_X x_2 \Leftrightarrow f(x_1)\prec_Y f(x_2)$.
\smallskip}

{\narrower\smallskip\noindent
{\bf Theorem 5:}
  In order that a poset $X$ be tonomorphic to the closed unit interval
  $[0,1] \subseteq \Reals$, it is necessary and sufficient that

\item{(i)}{$X$ be linearly ordered with both a minimum and a maximum element
(denoted $0,1$ respectively),}
\item{(ii)}{$X$ have a countable order-dense subset, and}
\item{(iii)}{every partition of $X$ into disjoint subsets $A\prec B$ be 
either of the form $A=\lla 0,x \rra$, $B=\la  x,1 \rra$ or of the form 
$A=\lla 0,x  \ra$, $B=\lla x,1 \rra$.}
\smallskip}

\noindent
Condition (iii) is just the requirement that both $A$ and $B$ be
intervals, as defined above, with one of them half-open and the other one
closed.  Its failure is often expressed by saying that $X$ has
either ``jumps''
(which occur when $A=\lla 0,a\rra$ and $B=\lla b,1\rra$) or ``gaps''
(which occur when $A$ lacks a supremum and $B$ lacks an infimum).  

The following result gives conditions for the isomorphism of Theorem 5 to
be topological as well as order-theoretic.

{\narrower\smallskip\noindent
{\bf Theorem 6:}
  Let $\Gamma$ be a set provided with both a linear order and a topology 
  such that

\item{$\bullet$}{with respect to the topology 
it is compact and connected and contains a countable dense subset,} 
\item{$\bullet$}{with respect to the order 
it has both a minimum and a maximum element, and}
\item{$\bullet$}{(with respect to both) 
it has the property that $\lla x,y\rra$ is topologically 
closed $\forall x,y\in \Gamma$.}

\noindent
  Then $\Gamma$ is isomorphic to the interval $[0,1]\subseteq\Reals$ by a
  simultaneous order and topological isomorphism.
\smallskip}

\vskip 2 true cm
\noindent
\cent{\bf 3. Causal Analysis in Terms of the Relation $K^+$}

\noindent
In this section we set forth the elements of a framework for doing global
causal analysis with $C^0$ metrics.  In particular, we introduce the
relation $K^+$, we define global hyperbolicity and causal curve, and we give a
simple ``intrinsic'' characterization of the latter in Theorem 20.  We
begin with some definitions, most of which are standard in the literature,
except that here they pertain to general $C^0$ metrics.

{\narrower\smallskip\noindent
{\bf Definitions 7:}
Let ${\cal M}$ be a $C^0$-lorentzian manifold with metric $g_{ab}$, and 
let $u^a$ be any vector field defining its time orientation. 
A timelike or lightlike
 vector $v^a$ is {\it future-pointing} if $g_{ab}v^a u^b < 0$ 
and {\it past-pointing} if $g_{ab}v^au^b>0$.  
Now let $I=[0,1]\subseteq\Reals$.  
A {\it future-timelike path} in ${\cal M}$ is a piecewise $C^1$, continuous
function 
$\gamma:[0,1] \to {\cal M}$ 
whose tangent vector
$\gamma^a(t)=(d\gamma(t)/dt)^a$ is future-pointing timelike whenever it is
defined (in particular $\gamma$ possesses a future-pointing timelike
tangent vector almost everywhere); a {\it past-timelike} path is defined
dually. 
The image of a future- or past-timelike path is a {\it timelike curve}. 
Let $\O$ be an open subset of $\M$. 
If there is a future-timelike
curve in ${\cal O}$ from $p$ to $q$, we write $q\in I^+(p,{\cal O})$,
and we call  $I^+(p,{\cal O})$ the {\it chronological future} of
$p$ relative to  ${\cal O}$.  If $q\in I^+(p,{\cal O})$, we write
$p<_{\cal O}q$.
Past-timelike paths 
and curves, the sets $I^-(p,{\cal O})$, and the 
symbol
$>_{\cal O}$ are defined analogously.  If ${\cal O}={\cal M}$, we may omit
it, so
$I^+(p,{\cal M})=I^+(p)$, and so forth.  
\smallskip}

\noindent
Notice that a curve is defined herein to be automatically compact: it
will be convenient not to allow the domain of $\gamma$ to be an open (or
half-open) interval in $\Reals$.

When the manifold obeys the ``chronology condition'' that it contain no
closed timelike curves, the relation $<$ is a {\it partial order}.
Actually, it is a consequence of the definition of $I^+(p)$ that it does
not contain $p$ when the chronology condition holds; hence $<$ is a partial
order observing the irreflexive convention.

That our basic causality relation be an order will be crucial for results
like Theorem 20, but that attribute alone will not be enough.  It will
also be important that the order be {\it closed}, by which we mean that it
be {\it topologically} closed when regarded as a subset of ${\cal
M}\times{\cal M}$.  Unfortunately $I^+$ does not enjoy this property, and
so we need to complete it in some manner to a relation which we will call
$K^+ =\;
\prec$.

To see that $<$ is not closed on ${\cal M}$, observe that if it were
closed, then we would have $x<y$ for the limits of any convergent sequences
$x_n\rightarrow x$ and $y_n\rightarrow y$ such that $x_n<y_n$ for each $n$;
but for example, in 2-dimensional Minkowski space ${\bf M}^2$, there 
are (in terms of Galilean coordinates $(t,z)$) sequences $x_n=(0,0) 
\rightarrow x=(0,0)$ and $y_n = (1,1-{1\over n})\rightarrow y=(1,1)$, for 
which $x_n < y_n$ but $x \not< y$ (see Figure 1).

%
\vbox{%
\includegraphics{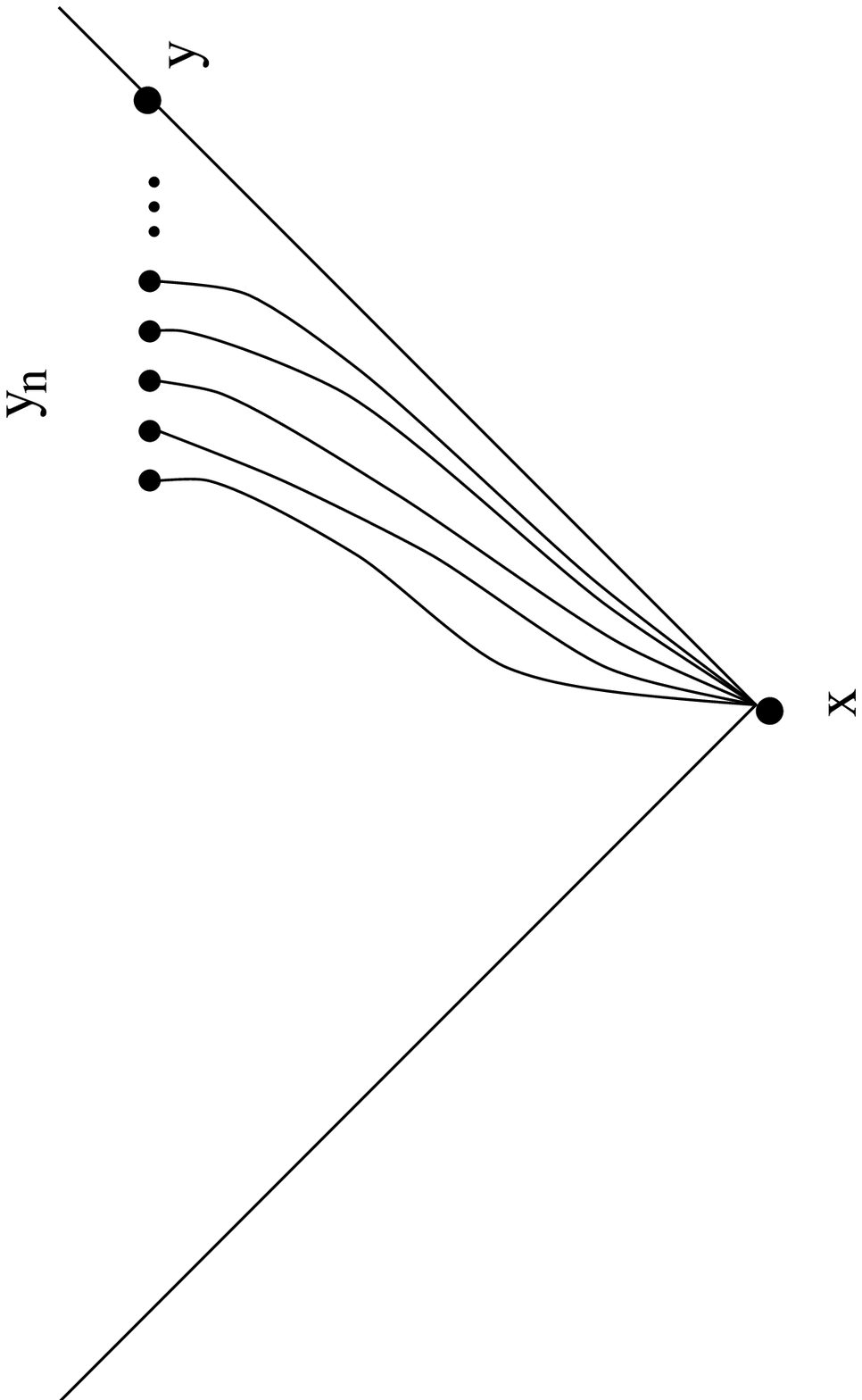}
\vskip 55 true mm
{\narrower\narrower\singlespace\eightrm\bigskip
\textfont0=\eightrm
\textfont1=\eighti \scriptfont1=\sixi
\textfont2=\eightsy \scriptfont2=\sixsy
\noindent
{\underbar {Figure~1:}}
Shown is a sequence of timelike curves in Minkowski space, sharing a common
past endpoint $x$. Their future endpoints $y_n$ approach 
the lightcone of $x$, and so 
$y = \lim y_n$ is not in $I^+(x)$,
illustrating the fact that the relation $I^+$ is not topologically 
closed. Here, $y$ is simply the future endpoint of a 
{\eightit null} curve emanating from $x$,
but in more general situations (as in the manifolds of Figures 2a--c, 
for example) there may be no causal curve from $x$ to 
$y$ at all.
\bigskip\bigskip}}

One possibility for constructing $K^+$ would just be to adjoin the
lightlike relations to $<$ (after all, $x$ and $y$ are lightlike related in
the above example), but that would not ensure closure in more general
situations.  Another possibility would be to simply replace $<$ by its {\it
closure} in ${\cal M}\times{\cal M}$, but that would not always be
transitive for every manifold on which one might want to define causal
curves.  It seems most straightforward, therefore, just to define $K^+ =\;
	\prec$ in terms of the properties we need (but {\it cf.} the second 
footnote in the Introduction):

{\narrower\smallskip\noindent
{\bf Definition 8:} 
  $K^+$ is the smallest relation containing $I^+$ 
  that is transitive and (topologically) closed.
\smallskip}

\noindent
That is, we define the relation
$K^+$, regarded as a subset of $\M\times\M$, to be the
intersection of all closed subsets $R\supseteq I^+$ with the property that
$(p,q)\in R$ and $(q,r)\in R$ implies $(p,r)\in R$.  (Such sets $R$ exist
because $\M\times\M$ is one of them.)  One can also describe $K^+$ as the
closed-transitive relation {\it generated by} $I^+$.  When $(p,q)\in K^+$,
we write $q\in K^+(p)$ or $p\prec q$ (read ``$p$ precedes $q$''), and 
we write $q\in K^+(p,\O)$ or $p\prec_{\O}q$, for the analogous
relation defined within $\O$. We also write $K(p,q)$ for the order-closed
interval $K^+(p)\cap K^-(q)\ideq\lla p,q \rra$. Note that $K^+(p,{\cal O})$
is always closed in $\O$, as an immediate consequence of the closure of
$\prec_{\O}$ itself. 

{\narrower\smallskip\noindent
{\bf Remark 8a:} 
In a Minkowski spacetime, $K^+$ agrees with the ordinary ``algebraic''
relation of causal precedence, for which $p$ precedes $q$ iff the vector
$q-p$ is future-timelike or future-lightlike.  
This follows immediately
from the fact that the ordinary relation is closed and transitive.  More
generally, $\prec_\O$ agrees with Minkowskian causal precedence for any
convex open subset $\O$ of a Minkowski spacetime. 
\smallskip}

Figures 2a--c illustrate properties of $K^+$ in some spacetimes that are  
locally Minkowski.

%
\vbox{%
\vskip 2 true cm
\includegraphics{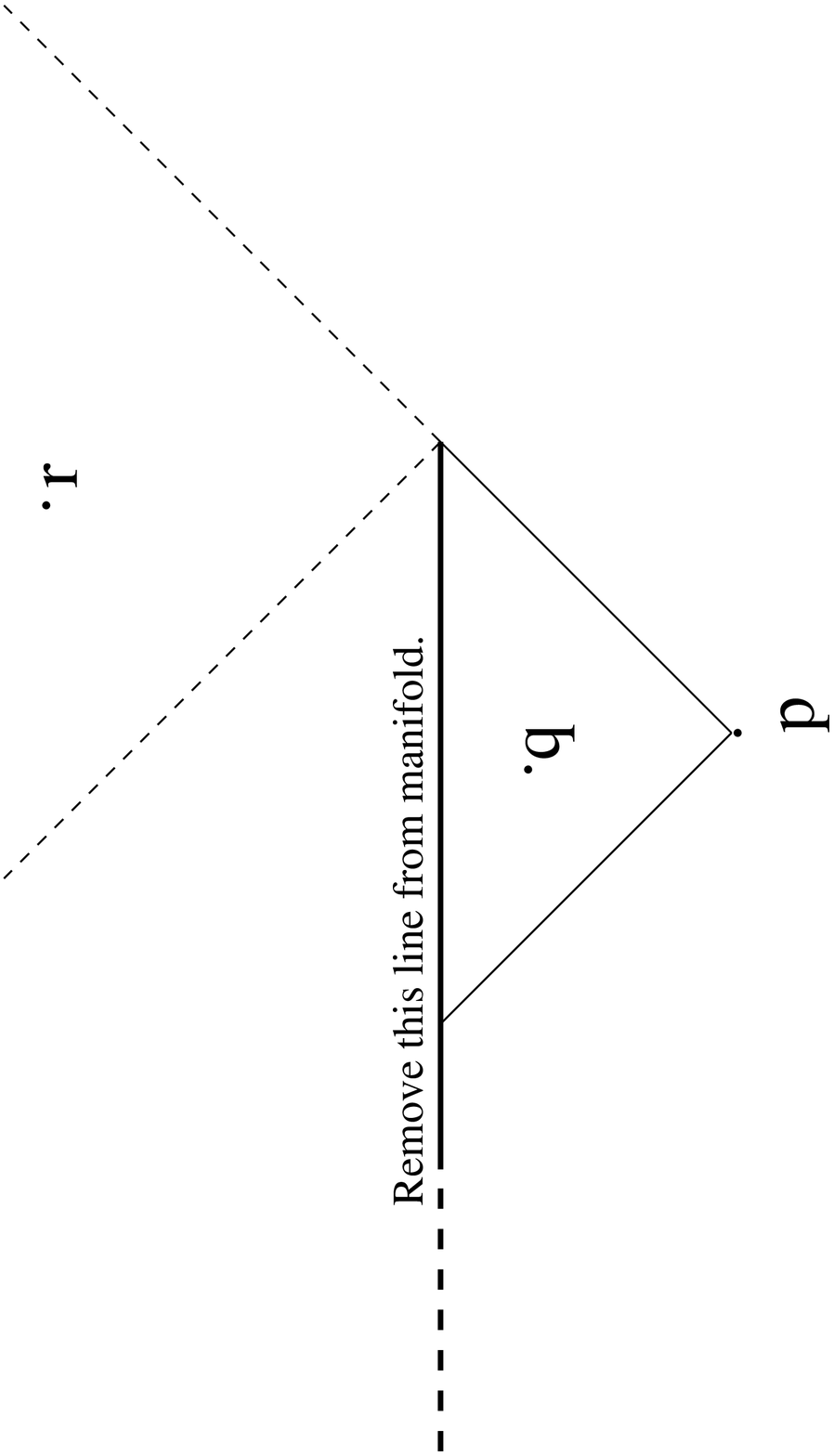}
\vskip 70 true mm
{\narrower\narrower\singlespace\eightrm\bigskip
\textfont0=\eightrm
\textfont1=\eighti \scriptfont1=\sixi
\textfont2=\eightsy \scriptfont2=\sixsy
\textfont\bffam=\eightbf
\noindent
{\underbar {Figure~2a:}}
Here we remove a closed half-line from 2-dimensional Minkowski space
${\bf M}^2$. We have $q\in I^+(p)\subseteq K^+(p)$.  The closure of 
$I^+(p)$, denoted ${\overline {I^+(p)}}$ and contained in $K^+(p)$, 
lies entirely below the removed half-line, and consists of every point 
between or on the thin solid lines. 
While $r$ is in $K^+(p)$, it is not in ${\overline {I^+(p)}}$; the same is 
true for any point between or on the dashed lines. Indeed, $r$ is in 
the interior of $K^+(p)$, although $p$ is on the boundary of $K^-(r)$ and
the pair $(p,r)$ itself is on the boundary of the relation $\prec$.
\bigskip\bigskip}}
%
%
\vbox{%
\includegraphics{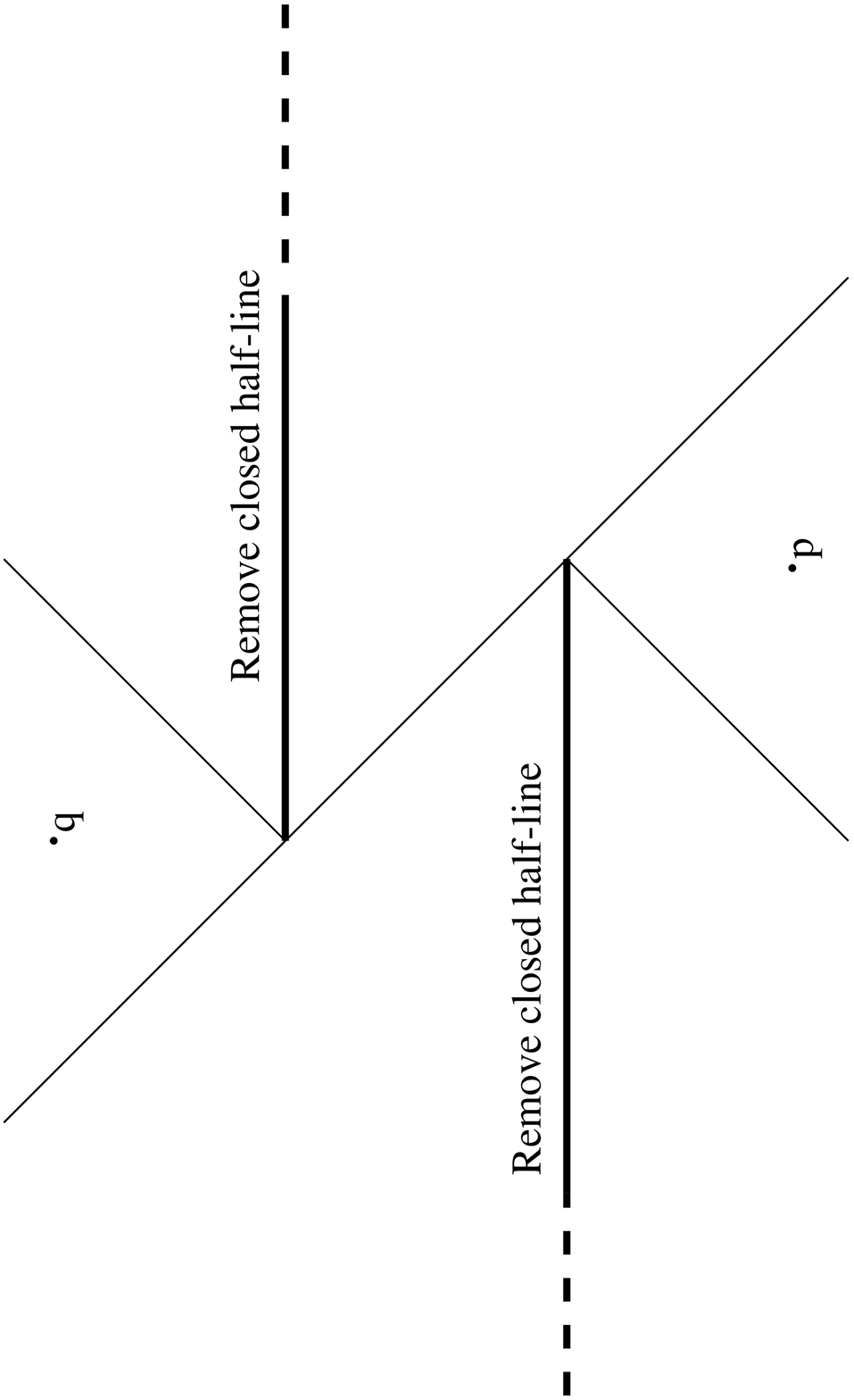}
\vskip 75 true mm
{\narrower\narrower\singlespace\eightrm\bigskip
\textfont0=\eightrm
\textfont1=\eighti \scriptfont1=\sixi
\textfont2=\eightsy \scriptfont2=\sixsy
\textfont\bffam=\eightbf
\noindent
{\underbar {Figure~2b:}}
Here we remove two closed half-lines from ${\bf M}^2$.  The slanted lines
are (incomplete) null geodesics.  The point $q$ is not in
${\overline{I^+(p)}}$ but it is in (the interior of) $K^+(p)$.  Indeed, the
pair $(p,q)$ is in the interior of the relation $\prec$.
\bigskip\bigskip}}
%

Although we have defined $\prec$ for a general manifold, its usefulness is
confined essentially to spacetimes (or their subsets) fulfilling the
asymmetry axiom (i) of Definition 3 above.  This motivates the following
definition, which we record here together with the definitions of two other
causal regularity conditions we will invoke repeatedly.

{\narrower\smallskip\noindent
{\bf Definition 9:}
  An open set ${\cal O}$ is $K$-{\it causal} iff the relation $\prec$ induces
  a (reflexive) partial ordering on ${\cal O}$; {\it i.e.} iff the asymmetry
  axiom, that $p\prec q$ and $q\prec p$ together imply $p=q$, holds for
  $p,q\in{\cal O}$. 
\smallskip}

{\narrower\smallskip\noindent
{\bf Definition 10:}
A subset of $\M$ is {\it $K$-convex} (also called {\it causally convex
with respect to $K^+$}) iff it includes the interval $K(p,q)$ between every
pair of its members, that is, iff it contains 
along with $p$ and $q$ 
any $r\in\M$ for which $p \prec r \prec q$.
\smallskip}  

{\narrower\smallskip\noindent
{\bf Definition 11:}
  A $K$-causal open set $\O\subseteq \M$ is {\it globally hyperbolic} iff,
  for every pair of points $p,q\in \O$, the interval $K(p,q)$ is compact
  and contained in $\O$. 
\smallskip}

\noindent
Notice that Definition 11 requires a globally hyperbolic set to be
$K$-convex.  Notice also that the above definitions have obvious analogues
for any other causality relation $R$, such as $R=I^+$.  In particular, a
set $S$ will be called $R$-convex iff 
$p,\, q\in S \implies R(p,q)\subseteq S$.

%
\vbox{%
\includegraphics{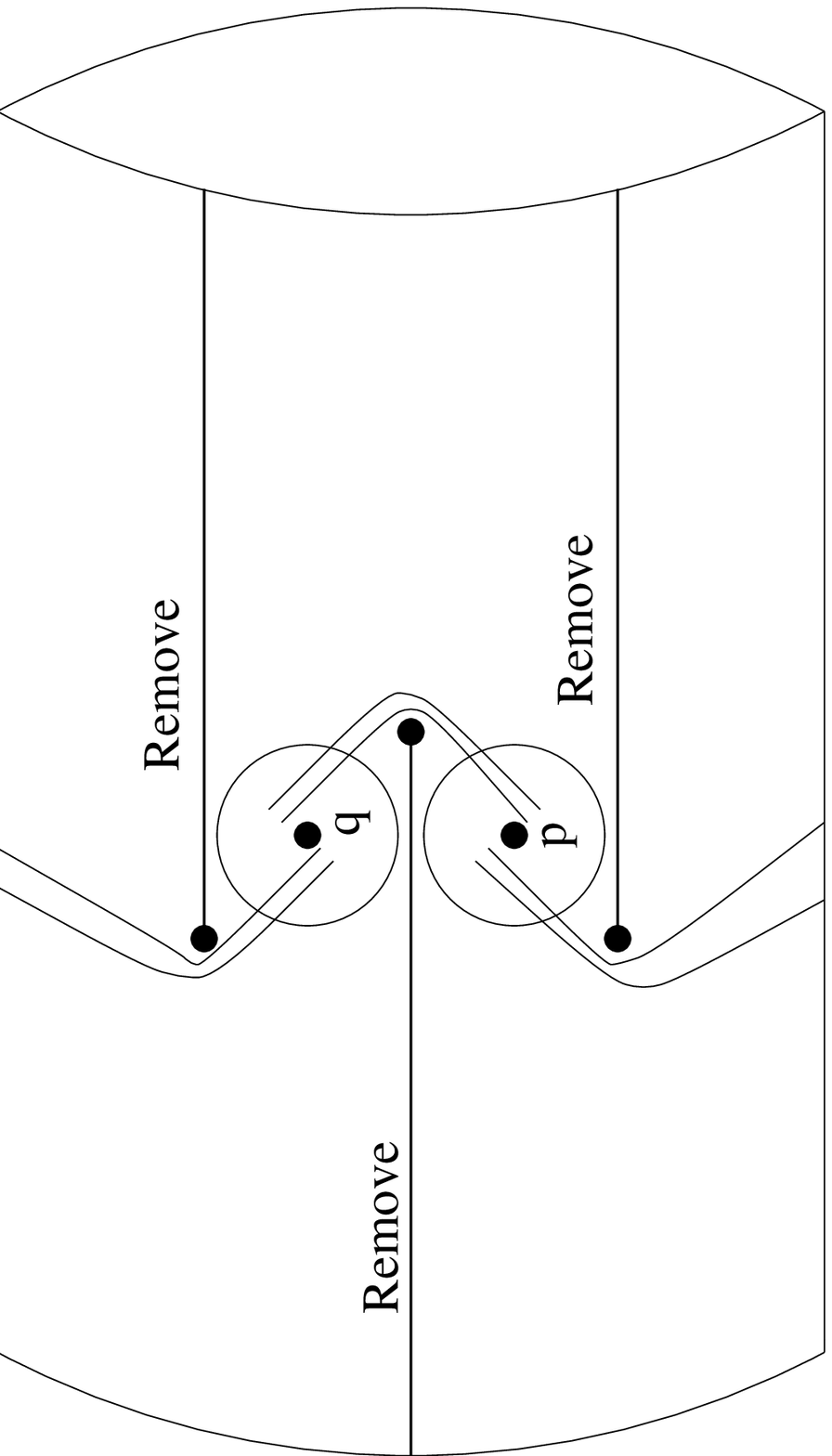}
\vskip 55 true mm
{\narrower\narrower\singlespace\eightrm\bigskip
\textfont0=\eightrm
\textfont1=\eighti \scriptfont1=\sixi
\textfont2=\eightsy \scriptfont2=\sixsy
\textfont\bffam=\eightbf
\noindent
{\underbar {Figure~2c:}}
Shown is a timelike cylinder derived from 
${\bf M}^2$
with Cartesian coordinates $(t,z)$ by periodically identifying $t$ (say
with period $4$, so $(-2,z)$ and $(2,z)$ denote the same point).  From this
spacetime, remove three closed half-lines, each parallel to the $z$-axis:
the half-lines starting at $(\pm 1,-1)$ and moving off to the right, and
the half-line starting from $(0,0)$ and moving off to the left.  The
resulting manifold has families of timelike curves as depicted, which
implies that $p\prec q \prec p$, even though no causal curve can actually
join $p$ to $q$.  This example also illustrates that causal convexity with
respect to $I^+$ need not imply causal convexity with respect to $K^+$.
\bigskip\bigskip}}
%

The next two lemmas relate $\prec_\U$ to $\prec_\O$ in a situation where
$\U\subseteq\O$.  The second of these lemmas will be needed only in
circumstances where it will be a trivial corollary of Lemma 15 below, but
we give it in full strength here, because its proof --- though a bit
lengthy --- is actually more elementary than the proof we have for Lemma
15.

{\narrower\smallskip\noindent
Let $\U$ and $\O$ be open subsets of $\M$ with $\U\subseteq\O$, and let 
$p, q \in\U$.  Then:
\smallskip
\noindent
{\bf Lemma 12:} 
  $p \prec_\U q$ implies  $p \prec_\O q$;
\smallskip
\noindent
{\bf Lemma 13:}
  $p\prec_\O q$ implies $p \prec_\U q$
  if $\U$ is causally convex with respect to $\prec_\O$.
\smallskip}

{\narrower\smallskip\noindent
{\bf Proofs:} 
Without loss of generality, we can take $\O=\M$, and we do so to ease the
notational burden on the proofs.  It is immediate from the definitions that
the {\it{}restriction} $\prec \! | \, \U$ of $\prec$ to $\U$ is closed and
transitive, and it also plainly contains the relation $<_{\U}$.  Therefore,
as $\prec_{\U}$ is the {\it smallest} relation with these properties, we
have $\prec_\U \,\subseteq \, \prec \! |\U$, which is the content of
Lemma 12.

Conversely, the content of Lemma 13 is the statement that
$\prec\!|\U \ \subseteq\ \prec_\U$, which, in view of Lemma 12, actually
means that they are equal.
To prove this, we will construct a relation $\preceq$ on ${\cal M}$ which
will obviously restrict to $\prec_{\cal U}$ on ${\cal U}$, and we will
prove that $\preceq\ =\ \prec$.  We define the relation $\preceq$ as
follows:

\item{(i)}{If $p,q\in {\cal U}$, then $p\preceq q \Leftrightarrow 
p\prec_{\cal U} q$.}
\item{(ii)}{If either or both of $p,q$ are not in ${\cal U}$, 
then $p\preceq q \Leftrightarrow p\prec q$.}

\noindent
Obviously $\preceq|{\cal U}=\prec_{\cal U}$ by (i).  Also, (i), (ii), and
Lemma 12 imply the inclusion $\preceq\ \subseteq\ \prec$.  Thus we need
only prove the reverse inclusion, which we will do by showing that
$\preceq$ is closed and transitive and contains $I^+$.

Closure of $\preceq$ is straightforward.
Let $p_n\to p$ and $q_n\to q$, with $p_n\preceq q_n$.  If both $p$ and
$q\in\U$, then the openness of $\U$ implies that eventually $p_n\in\U$,
$q_n\in\U$, whence $p_n\prec_\U q_n$ by (i), whence 
$p \prec_\U q$ since $\prec_\U$ is closed, whence $p \preceq q$,
again by (i).  If, on the contrary, one of $p,
q$ is not in $\U$, then by (ii) we need only show $p \prec q$, 
but this follows
immediately from the closure of $\prec$ since,
as observed just above, $p_n \preceq q_n$ always
implies $p_n \prec q_n$, whether or not $p_n,\, q_n \in \U$.

Next we check transitivity. We assume that $p\preceq r\preceq q$, and we
check that $p\preceq q$. If $p,r,q\in {\cal U}$, then transitivity
follows from transitivity of $\prec_{\cal U}$. If $p$ and $r$ are
in ${\cal U}$ and $q$ is not, then 
$p\preceq r \Rightarrow p\prec r$ as we know,
and $r\prec q$ by part (ii) of the 
definition of $\preceq$, whence $p\prec q$ by transitivity of $\prec$, 
whence $p\preceq q$ by again applying part (ii) of the
definition of $\preceq$; likewise if $r,q\in {\cal U}$ and 
$p\notin \U$. 
The potentially dangerous case of $p,q\in {\cal U}$, $r\notin {\cal U}$
does not arise, since $p\prec r\prec q \implies r \in \U$ by the causal
convexity of ${\cal U}$.  If no more than one of $p,r,q$ are in ${\cal U}$,
then $p\preceq r\preceq q$ implies $p\prec r\prec q$, which implies $p\prec
q$, which in turn implies $p\preceq q$, once again by part (ii) of the
definition of $\preceq$.

Lastly, we must show that $\preceq \; \supseteq \; < \;$.  So suppose there
exists a timelike curve $\gamma$ from $p$ to $q$.  (Recall that `$p<q$'
means `$q\in I^+(p)$'.)  Because $p<q\Rightarrow p\prec q \Rightarrow
p\preceq q$ when either $p$ or $q$ is not in ${\cal U}$, the only
non-trivial case occurs when $p,\, q\in\U$.  But then, since $\U$ is
causally convex, $\gamma$ must remain within it; otherwise there would
exist $r\notin\U$ such that $p<r<q$, whence we would have $p \prec r \prec
q$, contrary to causal convexity.  Therefore $q\in I^+(p,\U)$, {\it i.e.}
$p \prec_\U q$, whence $p \preceq q$ by (i).

Thus, $\preceq$ is closed and transitive, it contains $<$, and it is
contained within $\prec$. Since $\prec$ is defined to be the smallest
closed and transitive relation containing $<$, we have 
$\preceq \ =\  \prec$.
$\quad\dal$\smallskip}

Precisely because the relation of $K^+$ to curves is rather indirect,
certain lemmas which would be straightforward for $J^+$ are more difficult
to establish for $K^+$.  The next two results are of this nature, and we
resort to transfinite induction to prove them.  (Notice, incidentally, that
no assumption of $K$-causality is involved.)

{\narrower\smallskip\noindent
{\bf Lemma 14:} 
 Let $B$ be a subset of $\M$ with compact boundary, 
 and let $x \prec y$ with $x \in B$, $y \notin B$ (or vice versa).
 Then $\exists w \in \partial B$ such that $x \prec w \prec y$.
\smallskip}

{\narrower\smallskip\noindent
{\bf Proof:}
We can regard the relation $K^+$ as having been built up via transfinite
induction (see {\it e.g.} [6]) from $I^+$ by adding pairs
$x\prec y$ as required.  Specifically, we look at each stage of the
induction for a pair which is not yet present, but which is
implied by either transitivity or closure, and if we find one, we add it.
Since there are at most $2^{\aleph_0}$ possible pairs, the process must
terminate.

Let the steps in this process be labeled by the ordinal number $\alpha$,
with the corresponding relation as built up at stage $\alpha$ denoted by
$\prec^\alpha$.  We now make the inductive hypothesis $H_\alpha$ that, if
at stage $\alpha$ there are points $x \prec^\alpha y$ with $x\in B$,
$y\notin B$, then there exists $w\in\partial B$ such that $x\prec w\prec y$
(meaning that $w$ will lie between $x$ and $y$ in the {\it ultimate}
relation $\prec$).

Clearly $H_\alpha$ is valid for $\alpha=0$, so we need only check that
it holds for $\alpha$ assuming it holds for all $\beta < \alpha$.

Here there are two cases, depending on whether $\alpha$ is a limit ordinal
or not. 
If $\alpha$ is a limit ordinal, then $x \prec^\alpha y$ really just means
that $x \prec^\beta y$ for some $\beta<\alpha$, so  $H_\alpha$ follows
trivially. 

If $\alpha$ is not a limit ordinal, let $\alpha=\beta+1$, and observe that
we can assume that the relation $x \prec^\alpha y$ is precisely the one which
was added in passing from $\beta$ to $\alpha$. 
There are then two sub-cases depending on why this relation was added:

(i) If $x \prec^\alpha y$ was added by transitivity, then there must have
been a $z$ for which $x \prec^\beta z \prec^\beta y$. If $z\in B$, then
the inductive hypothesis $H_\beta$ implies 
the existence of $w\in\partial B$ such that $z \prec w \prec y$, whence 
$x \prec z \prec w \prec y \implies x \prec w \prec y$, as desired.  On the
other hand, if $z$ lies outside of $B$, then $H_\beta \implies \exists w\in
\partial B$ such that $x \prec w \prec z \prec y \implies x \prec w \prec
y$, again as desired.

(ii) If $x \prec^\alpha y$ was added because of the existence of
sequences $x_n\to x$, $y_n\to y$ with $x_n \prec^\beta y_n$ then $H_\beta
\implies \forall n\ \exists w_n\in\partial B$ 
such that $x_n \prec w_n \prec y_n$.   
(Strictly speaking, we can be sure that (eventually) $x_n\in B$, 
$y_n\notin B$ only if $x$ is in the interior of $B$ and $y$ is in the
interior of its complement; but in fact this is the only situation we need
consider, because the lemma itself holds trivially if either $x$ or $y$ is in 
$\partial B$.)
Then,
as $\partial B$ is compact, we can pass to a convergent subsequence of
the $w_n$ and assume therefore that $w_n\to w$ for some $w\in\partial B$.
{}From this and the closure of $K^+$ it follows directly that 
$x \prec w \prec y$, thus completing the proof.
$\quad\dal$\smallskip}

{\narrower\smallskip\noindent
{\bf Lemma 15:} 
  Let $B$ be any open set in $\M$ with compact closure.
  If there are $x, y \in B$ with $x \prec y$ but not $x \prec_B y$, 
  then $\exists z\in \partial B $ such that $x \prec z \prec y$.
\smallskip}

{\narrower\smallskip\noindent
{\bf Proof:}
We will employ a closely similar transfinite induction as before, with the
inductive hypothesis $H_\alpha$ being that if $x \prec^\alpha y$ but
$x\not\prec_B y$ then $\exists z \in\partial B$ such that $x \prec z \prec
y$.  For this proof we adopt the shorthand that $x \prec y$ is {\it
unnatural} (with respect to $B$) iff $x \not\prec_B y$.

Clearly $H_\alpha$ holds for $\alpha=0$ since $K^+_0=I^+$.  Now suppose
$H_\beta$ holds for all $\beta < \alpha$,
and let $x\prec^{\alpha} y$ be unnatural.
Clearly if $x \prec^\beta y$ for
some $\beta<\alpha$ then $H_\beta \implies\exists z\in \partial B$ such
that $x\prec z \prec y$, and we are done.  So, as before, the only
non-trivial case is where $x \prec^\alpha y$ was added at stage $\alpha$,
which means that we have $\alpha=\beta+1$ and $x\not\prec^\beta y$.  Then,
as before, there are two sub-cases.

In the sub-case where $x\prec^\alpha y$ was added for transitivity, there must
exist $q$ such that $x \prec^\beta q \prec^\beta y$.  If $q\notin B$
then, by the previous lemma, $x \prec q \implies \exists z\in\partial B$
such that  
$x \prec z \prec q \implies x \prec z \prec q \prec y \implies x \prec z
\prec y$, as required.  If, on the other hand $q\in B$, then 
(by the transitivity of $\prec_B$)
at least one of
$x \prec q$ or $q \prec y$ must be unnatural.  If  $x \prec q$ is unnatural
then 
$H_\beta\implies\exists z\in\partial B$ such that $x \prec z \prec q \implies 
x \prec z \prec q \prec y \implies x \prec z \prec y$;  if $q \prec y$ is
unnatural then we see similarly that 
$H_\beta\implies\exists z\in \partial B$ such that 
$q \prec z \prec y \implies x \prec q \prec z \prec y\implies x \prec z
\prec y$.

In the sub-case where $x\prec^\alpha y$ was added because there exist
sequences $x_n\to x$, $y_n\to y$ with $x_n \prec^\beta y_n$, we can
conclude that $x_n \prec y_n$ is unnatural for sufficiently great $n$
(otherwise we would have $x_n \prec_B y_n$ for arbitrarily great $n$,
whence $x \prec_B y$ by the topological closure of $\prec_B$).  Thus we can
assume without loss of generality that $x_n \prec y_n$ is unnatural for all
$n$.  The inductive hypothesis $H_\beta$ then provides us for all $n$ with
$z_n\in \partial B$ such that $x_n \prec z_n \prec y_n$, and by the
compactness of $\partial B$, we can pass to a subsequence such that the
$z_n$ converge to some $z$ in $\partial B$.  But then the closure of
$\prec$ implies that $x \prec z \prec y$ as required.
$\quad\dal$\smallskip}

The last two results allow us to prove that every $K$-causal manifold is
what might be called {\it locally $K$-convex} or ``strongly causal with
respect to $K^+$:''

{\narrower\smallskip\noindent
{\bf Lemma 16:} 
  If $\M$ is $K$-causal then every element of $\M$ possesses 
  arbitrarily small $K$-convex open neighborhoods ({\it i.e.} $\M$ is {\it
  locally $K$-convex}).
\smallskip}

{\narrower\smallskip\noindent
{\bf Proof:}
Given any neighborhood $N$ of $p$ in $\M$, we can find an open
sub-neighbor\-hood $B\ideq U_0$ of compact closure,
and within $B$ a nested family of open
sub-neighborhoods $\{U_n\}$ which shrink down to $p$ and which are all
causally convex with respect to $\prec_B$.  (For $B$ sufficiently small,
such a family can be obtained 
by taking neighborhoods which are causally convex with respect
to some flat metric within $B$ whose light cones are everywhere wider than
those of the true metric $g_{ab}$.  That this construction works follows
immediately from Remark 8a, the continuity of $g_{ab}$, 
and the definitions of $\prec_B$ and of causal
convexity.)

Now suppose the theorem fails at $p$.  Then for some neighborhood $N$ of
$p$, none of the $U_n$ just described can be causally convex with respect
to $\prec$ itself.  This means that for each $n$ there are 
$x_n, z_n \in U_n$ such that $x_n\prec y_n \prec z_n$ with 
$y_n \notin U_n$.  Now either $y_n$ belongs to $B$ or it doesn't.  
If it does, then (in the terminology of the proof of Lemma 15) either
$x_n \prec y_n$ or $y_n \prec z_n$ 
must be unnatural with respect to $B$
 (because $U_n$ is causally convex within $B$).  Suppose
that $x_n \prec y_n$ is unnatural (the other possibility being analogous).
Then Lemma 15 provides us with $w_n\in \partial B$ such that 
$x_n \prec w_n \prec y_n \prec z_n$, whence 
$x_n \prec w_n \prec z_n$.  On the other hand, if $y_n$ lies outside of $B$
then a precisely analogous application of Lemma 14 furnishes us, once
again, with a $w_n\in \partial B$ such that $x_n \prec w_n \prec z_n$.

In either case, we get a sequence $w_n$ of elements of the compact set
$\partial B$, which lie causally between $x_n$ and $z_n$ and, passing to a
subsequence if necessary, an element $w\in \partial B$ to which the $w_n$
converge.  But this would imply 
$p = \lim x_n \prec w \prec \lim z_n = p$, contradicting the
$K$-causality of $\M$.
$\quad\dal$\smallskip}

In the remainder of this section, we define the notion of causal curve
using $\prec$, and we derive a simple intrinsic characterization of this
notion which is valid in an arbitrary $K$-causal manifold.  Among the
lemmas leading up to this result, Lemma 19 seems of some interest
in itself as an independent simple characterization of causal curve.

{\narrower\smallskip\noindent
{\bf Definition 17:}
 A {\it causal curve} $\Gamma$ from $p$ to $q$ is the image of a $C^0$ map
 $\gamma:[0,1]\to{\cal M}$ with $p=\gamma(0)$, $q=\gamma(1)$, and such that
 for each $t\in (0,1)$ and each open $\O\ni\gamma(t)$, there
 is a positive number $\epsilon$ (depending on ${\cal O}$) such that
$$
\eqalignno
   {&t'\in (t,t+\epsilon)\Rightarrow \gamma(t)\prec_{\O}\gamma(t') 
    \quad,\; {\rm and}  &{\rm (i)}\cr
    &t'\in (t-\epsilon,t)\Rightarrow \gamma(t')\prec_{\O}\gamma(t)
    \quad .&{\rm (ii)}\cr}
$$ 
 Moreover, condition (i) above is required to hold for $t=0$, and condition 
 (ii) for $t=1$.  The point $p=\gamma(0)$ is called the {\it initial endpoint}
 of $\gamma$, and $q=\gamma(1)$ is called its {\it final endpoint}.
\smallskip}

\noindent
A causal curve is thus defined to be the image of what might be called a
locally increasing mapping, and rewording the definition accordingly
results in a slightly more compact formulation as follows. 

{\narrower\smallskip\noindent
 A path $\gamma$ is {\it causal} iff for every $t\in[0,1]$ and every open
 neighborhood $\O$ of $\gamma(t)$ there exists a neighborhood of $t$ in
 $[0,1]$ within which $t'< t'' \implies \gamma(t') \prec_\O \gamma(t'')$; a
 causal curve is the image of a causal path.
\smallskip}
\noindent
We may remark also that the requirement in Definition 17 that
$\gamma$ be $C^0$ is actually redundant, being implicit in
conditions (i) and (ii).

{\narrower\smallskip\noindent
{\bf Lemma 18:}
 If $\Gamma={\rm image}(\gamma)$ is a causal curve, then for every
 neighborhood ${\cal O}\supseteq\Gamma$ and every pair $t_i<t_f\in [0,1]$,
 we have $\gamma(t_i)\prec_{\cal O}\gamma(t_f)$.
\smallskip}

{\narrower\smallskip\noindent
{\bf Proof:}
Without loss of generality we can take $t_i=0$ and $t_f=1$; let us also set
$p=\gamma(0)$.  
Let 
      $t_0 = \sup \{t\in [0,1]\ |\ p \prec_\O \gamma(t) \}$.
Because
$\prec_\O$ is closed, $r := \gamma(t_0)$ also follows $p$
({\it i.e.} $p\prec_\O r$).  But then we must have $t_0=1$ since
otherwise, by Definition 17, 
there would exist $t_1 > t_0$ with 
$p \prec_\O r \prec_\O \gamma(t_1)$, whence $p \prec_\O \gamma(t_1)$ by 
transitivity, contradicting the definition of $t_0$.  $\quad\dal$ 
\smallskip}

Notice that the converse of Lemma 18 is in general false: 
$p\prec q$ need not imply the existence of a causal curve from $p$ to
$q$, even for $q$ in the interior of $K^+(p)$.
Notice also that, in the absence of $K$-causality,
Lemma 18 does not in general endow $\Gamma$ with a linear order,
because the asymmetry axiom can fail for $\prec$ restricted to $\Gamma$,
even if we replace $\prec$ by the intersection of the $\prec_\O$ for all
$\O\supseteq\Gamma$.

In a $K$-causal spacetime, Definition 17 of causal curve can be stated more
simply: a causal curve is the image of a continuous order-preserving map
from $[0,1]$ to $\M$.  We obtain this equivalence in the following lemma.

{\narrower\smallskip\noindent
{\bf Lemma 19:} 
  A subset $\Gamma$ of a $K$-causal spacetime is a causal curve if and only
  if it is  
 the image of a continuous increasing function $\gamma$ from $[0,1]$ to $\M$.
\smallskip}

\noindent
(By increasing, we mean that 
$t<t'\Rightarrow\gamma(t)\prec\gamma(t'),\ \forall t,t'\in[0,1]$.)

{\narrower\smallskip\noindent
{\bf Proof:} The forward implication (``only if'' clause) follows immediately 
from Lemma 18.  
To establish the reverse implication, we only need show that, for every
$t\in[0,1]$ and every open neighborhood $\O$ of $p=\gamma(t)$, there exists a
neighborhood of $t$ in $[0,1]$ within which
$t'<t''\implies\gamma(t') \prec_\O \gamma(t'')$ .  But given any such $\O$,
Lemma 
16 provides an open subset $\U$ of $\O$ which is causally convex and
contains $p$.  Then, since $\gamma$ is continuous, there exists a
neighborhood $[r,s]$ of $t$ in $[0,1]$ whose image by $\gamma$ lies within
$\U$; while for $r \le t' \le t'' \le s$ we have by assumption that
$\gamma(t')\prec\gamma(t'')$.   Hence 
$\gamma(t')\prec_\U\gamma(t'')$ by Lemma 13, which in turn entails
$\gamma(t')\prec_\O\gamma(t'')$, by Lemma 12.
$\quad \dal$\smallskip}

%

By the definition of $K$-causality, the function $\gamma$ of Lemma 19 can
be taken, without loss of generality, to be injective.  
Thus, the distinction between
a causal curve and a causal
path becomes essentially irrelevant in a $K$-causal spacetime; that is, a
causal curve $\Gamma$ becomes identifiable with the corresponding causal
path $\gamma:[0,1]\to\M$, up to reparameterization induced by a
diffeomorphism of $[0,1]$.

{\narrower\smallskip\noindent
{\bf Theorem 20:} 
  A subset $\Gamma$ of a $K$-causal manifold is a causal curve iff
  it is compact, connected and linearly ordered by $\prec \ = K^+$.
\smallskip}

{\narrower\smallskip\noindent
{\bf Proof:}
To prove the theorem, 
we will establish that
if $\Gamma\subseteq\M$ is compact, connected and linearly ordered by
$\prec$ then it is homeomorphic (by an order-preserving
correspondence) to the unit interval in $\Reals$.
The theorem will then be immediate from Lemma 19.

Let us apply Theorem 6.  To verify the hypotheses of that theorem, we need to
show that $\Gamma$ contains a countable dense subset, that it has minimum
and maximum elements, and that 
$\{z\in\Gamma | x \prec z \prec y\}$ is closed for all $x$ and $y$ in
$\Gamma$. 

The existence of a countable dense subset follows directly from the fact
that $\Gamma$ is a compact subset of a finite dimensional manifold.  (As
such, $\Gamma$ can be covered by a finite number of subsets of $\M$
homeomorphic to open balls in $\Reals^n$, therefore it is second-countable,
therefore it has a countable dense subset.)

The fact that $\Gamma$ contains minimum and maximum elements follows from
the circumstance that $\Gamma$ is compact and $\prec$ is closed.  This
implies that the set
$\Omega := \cap_{x\in\Gamma} (\Gamma \cap K^-(x))$ 
is non-empty (being the intersection of a nested family of non-empty closed
subsets of the compact set $\Gamma$).  But, since any element of $\Omega$
clearly precedes every element of $\Gamma$, we must have $\Omega = \{0\}$,
where $0$ is the desired minimum element.  Likewise, we obtain $1\in\Gamma$
such that $p\prec 1,\ \forall p\in\Gamma$.

That, finally, every interval in $\Gamma$ is topologically closed follows
immediately from the fact that the interval in $\Gamma$
bounded by $x,y\in\Gamma$ can be expressed as the intersection of three
closed sets, $K^+(x)\cap K^-(y)\cap\Gamma$.  (Recall that $K^\pm(x)$ is
closed since $\prec$ itself is.)  
$\quad\dal$\smallskip}

{\narrower\smallskip\noindent 
{\bf Remark:}
The proof shows that $\Gamma$ would be a curve, even without the condition
of $K$-causality; but this condition is not redundant because, in its
absence, one can find examples where $\Gamma$ turns out to be a spacelike
curve rather than a causal one. 
\smallskip}

\vskip 2 true cm
\noindent
\cent{\bf 4. A Bicompactness Theorem for  Spaces of Causal Curves}
\par\nobreak
\noindent
Theorem 23 of this section will be the main fruit of our work in this
paper.  Given the general characterization of a causal curve (in a
$K$-causal spacetime) as a compact connected subset linearly ordered by
$\prec$, the derivation of Theorem 23 is relatively simple and
straightforward.  All that is needed, basically, is to establish that the
limit of a sequence of causal curves is a causal curve, which is the
content of the next proposition.

{\narrower\smallskip\noindent 
{\bf Theorem 21:} 
  In a $K$-causal $C^0$-lorentzian manifold ${\cal M}$, 
  a compact Vietoris limit of a sequence (or net) of causal curves
   is also a causal curve.
\smallskip}

{\narrower\smallskip\noindent
{\bf Proof:}
Let us verify the conditions of Theorem 20 for the limit $\Gamma$ of a
sequence of causal curves $\Gamma_n$.

In order to establish that $\Gamma$ is linearly ordered by $\prec$, let $p$
and $q$ be points on $\Gamma$, and let $\O_1\ni p$ and $\O_2\ni q$ be open
neighborhoods.  Then $\Gamma\in{\cal B}(\M;{\cal O}_1,{\cal O}_2)$,
together with $\Gamma_n \to \Gamma$, implies 
$\Gamma_n \in{\cal B}({\M};{\cal O}_1,{\cal O}_2)$ for $n$ sufficiently
great.  By letting $\O_1$ shrink down around $p$ and $\O_2$ shrink down
around $q$, we obtain sequences of points $p_n, q_n \in \Gamma_n$ such that
$p_n\to p$ and $q_n\to q$.  Since the $\Gamma_n$ are causal curves, Lemma
19 tells us that, for each $n$, $p_n\prec q_n$ or vice versa.  Hence one of
these holds for an infinite number of $n$, and the fact that $\prec$ is
closed then implies that $p\prec q$ or vice versa.  Thus every two points
of $\Gamma$ are related by $\prec$, as required.

That $\Gamma$ is compact is immediate from the first part of Lemma 2.  That
it is  connected follows from the second part, in virtue of the fact that
$\Gamma$ has a compact neighborhood within which the $\Gamma_n$ must
ultimately lie.  (Such a neighborhood exists because $\Gamma$, being
compact, can be covered by a finite number of open sets of compact
closure.  The union $\O$ of these sets is then an open neighborhood of
$\Gamma$ which ultimately contains the $\Gamma_n$ since ultimately 
$\Gamma_n\in\B(\O;\M)$.  Its closure $\bar\O$ is the desired compact
neighborhood of $\Gamma$.)
$\quad \dal$\smallskip}

\noindent
We will also use the following, unsurprising lemma.

{\narrower\smallskip\noindent 
{\bf Lemma 22:} 
In a $K$-causal manifold, let the causal curve $\Gamma$ be the Vietoris
limit of a sequence (or net) of causal curves $\Gamma_n$ with initial
endpoints $p_n$ and final endpoints $q_n$.  Then the $p_n$ converge to the
initial endpoint of $\Gamma$ and the $q_n$ to its final endpoint.
\smallskip}

{\narrower\smallskip\noindent
{\bf Proof:}
Since $\Gamma$ is compact, it has a compact neighborhood $\N$ within
which the $\Gamma_n$ eventually lie; hence we can assume without loss of
generality that there exists $p\in\N$ to which the $p_n$ converge.  In
fact, $p$ must belong to $\Gamma$, as follows from the more general fact
that 
{\it $\Gamma$ contains any point $x$ at which the $\Gamma_n$ accumulate}
in
the sense that they eventually meet any neighborhood of $x$.  (Proof: If
some point $x$ is not in $\Gamma$, then, as $\Gamma$ is compact, $x$
possesses a closed neighborhood $\K$ disjoint from $\Gamma$.  The
complement of $\K$ is then a neighborhood of $\Gamma$, and therefore must
eventually include all the $\Gamma_n$.  Hence the latter do not accumulate
at $x$, being disjoint from its neighborhood $\K$.)  Finally, it is easy to
see that $p$ must be the minimum element of $\Gamma$.  Indeed, any
$r\in\Gamma$ is the limit of some sequence of elements $r_n\in\Gamma_n$ (as
follows from the fact that every neighborhood ${\cal O} \ni r$ must
eventually meet all of the $\Gamma_n$), and this, together with the fact
that $p_n{}\prec{}r_n,\forall{n}$ (by Theorem 20), implies that $p \prec
r$.  Thus $p$ must be the initial endpoint of $\Gamma$ according to Theorem
20.  The statement for $q$ follows dually.  $\quad \dal$\smallskip}

\noindent
With this technicality out of the way, Theorem 21 implies:

{\narrower\smallskip\noindent
{\bf Theorem 23:}
Let $\O$ be a globally hyperbolic open subset of a $C^0$-lorentzian
manifold $\M$, and let $\P$ and $\Q$ be compact subsets of $\O$.  Then the
space of causal curves from ${\cal P}$ to ${\cal Q}$ is bicompact.
\smallskip}

{\narrower\smallskip\noindent 
{\bf Proof:} 
Let $\Gamma_n$ be a net of causal curves from ${\cal P}$ to ${\cal Q}$.
Since ${\cal P}$ and ${\cal Q}$ are compact, the initial endpoints $p_n$
and final endpoints $q_n$ accumulate at some points $p\in{\cal P}$ and
$q\in{\cal Q}$.  Since $\O$ is open, we can find $p' \in I^-(p) \cap \O$
and $q' \in I^+(q)\cap\O$.  Set ${\cal K} := K(p',q')$; it is compact by
Definition 11 of global hyperbolicity.  We can assume without loss of
generality that all the $p_n$ belong to $I^+(p')\subseteq K^+(p')$ and the
$q_n$ to $I^-(q')\subseteq K^-(q')$, hence that all the $\Gamma_n$ lie in
${\cal K}$.  Since ${\cal K}$ is compact, Theorem 1 provides a subnet
$\Gamma_m$ converging to some compact set $\Gamma\subseteq{\cal K}$, which
Theorem 21 assures us is actually a causal curve.  Finally, Lemma 22
assures us
(in view of the circumstance that $\O$ is $K$-causal by the definition of
global hyperbolicity) that $p$ is in fact the initial endpoint of
$\Gamma$ and $q$ is its final endpoint; {\it i.e.} $\Gamma$ is a causal
curve from $p$ to $q$.  Thus, the space of causal curves from ${\cal P}$ to
${\cal Q}$ is compact.

Additionally, since ${\cal K}$ is compact Hausdorff, the Vietoris topology on 
$2^{\cal K}$ is also Hausdorff, again by Theorem 1.  Thus, the space of 
causal curves from ${\cal P}$ to ${\cal Q}$ is Hausdorff as well.
$\quad\dal$\smallskip}

One can also prove, conversely to Theorem 23, that compactness of the space
of causal curves between arbitrary pairs of points $p,q$ implies global
hyperbolicity.  Similarly one can readily prove generalized forms of many
other results which are familiar in the $C^2$ setting, for example it is
not hard to see that in a globally hyperbolic manifold, $K^+ = J^+ = \bar
{I^+}$, where $J^+(p)$ is defined as the union of all causal curves
emanating from $p$ ({\it cf.} Lemma 25 in Appendix B).

\vskip 2 true cm
\cent{\bf Acknowledgments}
\par\nobreak
\noindent
RDS would like to acknowledge partial support from the
National Science Foundation, grant NSF PHY 9307570, and from Syracuse
University Research Funds.  EW would like to thank E. Tymchatin for a
discussion concerning Theorem 6.

\vskip 2 true cm
\cent{\bf References}
\par\nobreak

\item{[1]}
  Penrose, R., Sorkin, R.D., and Woolgar, E.,
  ``A Positive Mass Theorem Based on the Focusing and Retardation of
     Null Geodesics'', preprint 
    (available from the gr-qc archive as Paper: gr-qc/9301015);
 Penrose, R., {\it Twistor Newsletter} {\underbar {30}}, 1 (1990); 
 Sorkin, R.D., and Woolgar, E., {\it Proc. Fourth Can. Conf. on Gen.
  Rel. and Rel. Astrophys.}, ed. Kunstatter, G., Vincent, D.E., and Williams, 
  J.G., p. 206 (World Scientific, Singapore, 1992); {\it Proc. Sixth Marcel
  Grossmann Meeting on Gel. Rel.}, ed. Sato, H., and Nakamura, T., p. 754
  (World Scientific, Singapore, 1992).

\item{[2]} 
{Sorkin, R.D., ``Consequences of Spacetime Topology'', 
     in {\it Proceedings of the Third Canadian Conference on General
          Relativity and Relativistic Astrophysics}, 
           (Victoria, Can\-ada, May 1989), 
             ed. Coley, A., Cooperstock, F., and Tupper, B., 137-163
               (World Scientific, Singapore, 1990);
 Borde, A.,and  Sorkin, R.D.,
   ``Causal Cobordism: Topology  Change without Causal Anomalies'',
     (in preparation).}


\item{[3]}
 {Beem, J.K., and Ehrlich, P.E., {\it Global Lorentzian Geometry}
 (Marcel Dekker, New York, 1981).}

\item{[4]}
  {Nityananda, R., and Samuel, J.,
  {\it Phys. Rev.} D{\bf {45}}, 3862 (1992).}

\item{[5]} 
 {Klein, E., and Thompson, A.C., 
   {\it Theory of Correspondences},
   Can. Math. Soc. Ser. of Monographs and Advanced Texts 
   (Wiley, New York, 1984).}

\item{[6]}
{Kelley, J.L., {\it General Topology } (van Nostrand, Toronto, 1955).}

\item{[7]} 
 {Hawking, S.W., and Ellis, G.F.R., {\it The Large Scale Structure of
  Spacetime} (Cambridge, 1973).} 

\item{[8]}{Penrose, R., {\it Techniques of Differential Topology in
Relativity}, AMS Colloquium Publications, (SIAM, Philadelphia, 1972).}


\item{[9]}{Kamke, E., {\it Theory of Sets} (translated by
F. Bagemihl), (Dover, New York, 1950), Chapter III.}

\item{[10]}
{Blumenthal, L.M., and Menger, K.,
 {\it Studies in Geometry},
(W.H. Freeman, San Francisco, 1970),
 Theorem 12.3.}

\item{[11]}{Low, R., private communication (1995).}


\vskip 2 true cm
\cent{\bf Appendix A: Proofs of the Theorems in Section 2}
\noindent
Herein, for completeness, we prove the general mathematical theorems which
were collected in Section 2.  We deal with the Vietoris topology first, and
then the theorems concerning ordered sets.

In preparation for the  proof of Theorem 1, we need the following lemma
(``Alexander's Theorem'').

{\narrower\smallskip\noindent
{\bf Lemma 24:}
The topological space $\X$ is compact iff
every cover of $\X$ by open sets belonging to some fixed
{\it sub-base} for its
topology has a finite subcover.
\smallskip}

{\narrower\smallskip\noindent
{\bf Proof:}
The ``only if'' clause being trivial, let us assume that every cover by
sub-basic sets has a finite subcover and prove that then any net $x_n$ in
$\X$ has an accumulation point $x$.  Suppose there were no accumulation
point.  Then for each $x\in\X$ there would exist a sub-basic open set
$\O\ni x$ from which $x_n$ would eventually be absent.  It follows
immediately that $x_n$ would then have to be eventually absent from any
finite union of these $\O$s, whence no such union could include all of $\X$
(since each $x_n$ has to be {\it somewhere}).  Thus it would require an
infinite number of the sub-basic sets $\O$ to cover $\X$, contrary to
hypothesis.  $\quad\dal$\smallskip}

\noindent
Using this lemma we can prove the compactness of $2^\X$ very simply,
following [5].

{\narrower\smallskip\noindent
{\bf Theorem 1:}
  If the topological space $\X$ is bicompact, then so also is the space
  $2^\X$ of non-empty closed subsets of $\X$.
\smallskip}

{\narrower\smallskip\noindent
{\bf Proof:}
By definition, the sets of the form $\B(\O;\X)$ and $\B(\X;\O)$ provide a
sub-base for the Vietoris topology.  Then let
$$
  2^\X = \bigcup\limits_{i\in I} \B(\X;\O_i) \  \cup \   
         \bigcup\limits_{k\in K} \B(\O_k;\X)              \eqno (1)
$$
be a cover of $2^\X$ by such sets, where $i$ and $k$ range over distinct index
sets $I$ and $K$ (possibly empty).  We claim that the $\O_i$ together with
at most one of the $\O_k$ provide an open cover of $\X$ itself.  In fact if
$$
     \Y :=   \X  \backslash \bigcup\limits_{i\in I} \O_i  
$$
is not empty then (being closed) it is an element of $2^\X$, hence
contained in one of the $\B(\O_k;\X)$, say $\Y \in \B(\O_0;\X)$.  (This
follows from (1) since $\Y$ is not in any of the $\B(\X;\O_i)$,
being disjoint from all of the $\O_i$ by construction.)  But this means
precisely that $\Y\subseteq\O_0$, whence $\O_0$ combines with the $\O_i$ to
cover $\X$, as alleged.

Now since $\X$ is compact by assumption, we can find a finite subset of the
$\O_i$ which, together with $\O_0$, cover $\X$.  Letting $I_0\subseteq I$
be the corresponding finite set of indices, we have then
$$
     \X =  \bigcup\limits_{i\in I_0} \O_i \  \cup\   \O_0  \eqno(2)
$$
(where again, either $\O_0$ or the $\O_i$ might be absent because one of the
indexing sets $I$ or $K$ might be empty).  

We claim that the corresponding open sets $\B(\O_0;\X)$, and $\B(\X;\O_i)$
for $i\in I_0$, provide the desired finite subcover of (1).  In
fact, let $\S \in 2^\X$ be any closed subset of $\X$.  Either there is an
$i\in I_0$ such that $\S$ meets $\O_i$, or there is not.  In the former 
case, $\S$ is by definition an element of $\B(\X;\O_i)$, while in the
latter case, we see from (2) that
$\S$ must be a subset of $\O_0$, whence by definition an element of
$\B(\O_0;\X)$.  Thus 
$$
  2^\X = \bigcup\limits_{i\in I_0} \B(\X;\O_i) \  \cup \  \B(\O_0;\X),
$$
and we have shown that $2^\X$ is compact.

Lastly, we note that, since $\X$ is bicompact, it is regular, whence
$2^\X$ is Hausdorff by an easily proved property of the Vietoris topology
(Prop. 2.2.3. of [5]). 
$\quad\dal$\smallskip}

{\narrower\smallskip\noindent
{\bf Lemma 2:}

\item{(i)}{In a compact space every Vietoris limit is a compact set.}
\item{(ii)}{In a bicompact space, a Vietoris limit of connected sets is
connected.}

\smallskip}

{\narrower\smallskip\noindent 
{\bf Proof:} 
The first statement is trivial since any closed subset of a compact set is
compact.  For the second, observe that, if the limit $\Gamma$ of some
net $\Gamma_n$ were not connected, then it would be a disjoint
union\footnote{$^\dagger$}
{We use $\sqcup$ to denote disjoint union.}
$\Gamma = \Gamma' \sqcup \Gamma''$, where both $\Gamma'$  and $\Gamma''$ 
would be closed in $\Gamma$, and therefore compact, since $\Gamma$ itself
is compact by (i).  Because $\X$ is Hausdorff, $\Gamma'$  and
$\Gamma''$ would then possess disjoint open neighborhoods, 
${\cal O}_1$ and ${\cal O}_2$, in $\X$.  But that would mean that
$\Gamma\in{\cal B}({\cal O}_1\cup{\cal O}_2;{\cal O}_1,{\cal O}_2)$, 
whence, by the definition of the Vietoris topology, $\Gamma_n$ would also
belong to ${\cal B}({\cal O}_1\cup{\cal O}_2;{\cal O}_1, {\cal O}_2)$ for
sufficiently large $n$.  But then for such $n$, $\Gamma_n$ itself would not
be connected, contrary to hypothesis.
$\quad\dal$\smallskip}

  Next we turn to ordered spaces.  For the following theorem, see {\it e.g.}
reference [9].

{\narrower\smallskip\noindent
{\bf Theorem 5:}
   In order that a poset $X$ be tonomorphic ({\it i.e.} order-isomorphic)
   to the closed unit interval $I = [0,1] \subseteq \Reals$, it is
   necessary and sufficient that 

\item{(i)}{$X$ be linearly ordered with both a minimum and a maximum element
(denoted $0,1$ respectively),}
\item{(ii)}{$X$ have a countable order-dense subset, and}
\item{(iii)}{every partition of $X$ into disjoint subsets $A\prec B$ be 
either of the form $A=\lla 0,x \rra$, $B=\la  x,1 \rra$ or of the form 
$A=\lla 0,x  \ra$, $B=\lla x,1 \rra$.}

\smallskip}

{\narrower\smallskip\noindent
{\bf Proof:}
Condition (ii) provides us with a countable order-dense subset
$C_0\subseteq X$ which we clearly may assume to contain the $0$ and $1$
elements of $X$.  Similarly, the rational numbers between 0 and 1
(inclusive) are a countable order-dense subset $Q_0\subseteq I = [0,1]$.
We will first set up a bijection $f: C_0 \to Q_0$ and then complete it to
obtain an isomorphism between $X$ and $I$.

We will build up $f$ inductively by alternately choosing elements of $f$
and $f^{-1}$.  To begin with, we may assume that both $C_0$ and $Q_0$ (being
countable) are presented explicitly as sequences or {\it lists}
$C = (c_0, c_1, c_2, \dots) = (0, 1, \dots)$ and 
$Q = (q_0, q_1, q_2, \dots) = (0, 1, \dots)$,
where in each list, the order of the elements is immaterial except for the
initial two.  (No confusion should result from the fact that $0$ and $1$ are
being used to denote both real numbers and elements of $X$.)  We will 
obtain $f$
by inductively rearranging $C$
and $Q$ into lists
${\tilde C} 
   = ({\tilde c}_0, {\tilde c}_1, {\tilde c}_2, \dots) = (0, 1, \dots)$ and 
${\tilde Q} 
   = ({\tilde q}_0, {\tilde q}_1, {\tilde q}_2, \dots) = (0, 1, \dots)$,
such that the resulting induced correspondence 
$f({\tilde c}_i) = {\tilde q}_i$ $(i = 0,1,2,\dots)$ 
is tonomorphic from  $C_0 \subseteq  X$ to 
$Q_0 \subseteq [0,1] \subseteq \Reals$.

We may express this rearrangement algorithmically as a process in which we
begin 
with empty ${\tilde C}$ and ${\tilde Q}$ and gradually build them up
iteratively by {\it transferring} elements from $C$ and $Q$ until the
latter are exhausted, taking care that at each stage, the resulting
partially defined $f$ is maintained as a tonomorphism.  To do this we
simply repeat the following two-step ``loop'' a countable number of times.

\noindent
{\bf Step 1:} 
Remove the initial remaining element $c$ from $C$ and place it at the end
of ${\tilde C}$.  Then transfer from $Q$ to the end of ${\tilde Q}$ any
element $q$ (say the earliest one in the list) which will make $f$ remain a
tonomorphism.  Such a $q$ certainly exists because $Q$ is order-dense in
$I$ and the new element $c$ is situated (with respect to the order of
$X$) between a well-defined pair of the elements occurring earlier in
the list ${\tilde C}$.  

\noindent
{\bf Step 2:} 
This is the mirror image of Step 1,
with the roles of $C$ and $Q$ interchanged.
Remove the initial remaining element $q$ from $Q$ and place it at the end
of ${\tilde Q}$.  Then transfer from $C$ to the end of ${\tilde C}$ any
element $c$ (say the earliest one in the list) which will make $f$ remain a
tonomorphism.  Such a $c$ certainly exists because $C$ is order-dense in
$X$ and the new element $q$ is situated (with respect to the order of
$\Reals$) between a well-defined pair of the elements occurring earlier in
the list ${\tilde Q}$.  

After looping through these two steps a countable number of times, we will
have emptied both original lists and replaced them with re-ordered lists
${\tilde C} = (0, 1, {\tilde c}_2, {\tilde c}_3, \ldots)$ and 
${\tilde Q} = (0, 1, {\tilde q}_2, {\tilde q}_3, \ldots)$, 
thereby obtaining a densely defined, order-preserving mapping from $X$ to
$I$, or more precisely, an order-preserving bijection $f:C_0 \to Q_0$.  To
complete the proof we need to extend  $f$ to all of $X$.

Therefore, consider any $x\in X\backslash C_0$.  Such an $x$ partitions
$C_0$ into two sets $A$ and $B$ such that $A<x<B$ (taking $<$ to be the
order relation on $X$); this decomposition is called a {\it cut}.  The
corresponding cut in $Q$ determines a unique real number $r$, and we set
$f(x)=r$.  We claim that the resulting extended $f$ is an isomorphism.

To confirm this, let us first verify injectivity.  Consider
$x'<x''$.  Since $C_0$ is dense, there is an $x\in C_0$ such that $x'<x<x''$
(unless $\la x',x''\ra$ were empty, but that would mean that $X$ admitted a
decomposition 
$X=  \lla 0,x' \rra \sqcup  \lla x'', 1 \rra$
contrary to hypothesis).  Therefore, $x'$ and $x''$ induce unequal cuts in
$X$, since the cut due to $x'$ will have $x$ in the superior set of the
cut, while the $x''$ cut places $x$ in its inferior set.  But distinct cuts
in $X$ correspond to distinct cuts in $\Reals$; the cut defined by $f(x')$
will have $f(x)$ in its superior set while that defined by $f(x'')$ will
have $f(x)$ in its inferior set.  Therefore, $f$ is injective.  Notice that
we have also proved
$f(x') < f(x) < f(x'')$, whence $f(x')<f(x'')$, and our extended $f$ is
also order-preserving.

To prove surjectivity, we need only ask if all the irrationals in $I=[0,1]$
are in the image of $f$, since it is clear that the rationals in $[0,1]$
are.  Now any irrational $r\in [0,1]$ makes a cut in $Q$, say into $(R,S)$
with $R<S$. The inverse images by $f$  $R$ and $S$ give a
cut $(A,B)$ in $C_0\subseteq X$.  If there were no $x\in X$ inducing this
cut, then, setting 
${\bar A} = \{x\in X| (\exists a\in A)(x<a)\}$ and
${\bar B} = \{x\in X| (\exists b\in B)(b<x)\}$,
we would obtain a partition $({\bar A},{\bar B})$ of $X$ which would violate
condition (iii) of the theorem.   Thus, there is an $x\in X$ for
every irrational $r\in [0,1]$, so $f$ is surjective.

Lastly, we must verify that $x<x'\Leftrightarrow f(x)<f(x')$.  But
`$\Rightarrow$' has already been established above, while, in the present
setting, `$\Leftarrow$' follows from 
`$\Rightarrow$' since, by hypothesis, any two elements of $X$ are
comparable.  Thus, $f:X\to[0,1]\subseteq\Reals$ is a tonomorphism.
$\quad\dal$\smallskip}

Finally, we use the result just derived in the proof of the following
theorem ({\it cf}. [10]), which provides a sufficient condition
for an ordered topological space to be isomorphic to $[0,1]$.

{\narrower\smallskip\noindent
{\bf Theorem 6:}
  Let $\Gamma$ be a set provided with both a linear order and a topology such
  that

\item{$\bullet$}{with respect to the topology it is compact and connected and
      contains a countable dense subset,}
 \item{$\bullet$}{with respect to the order it has both a minimum and a
       maximum element, and}  
 \item{$\bullet$}{(with respect to both) it has the property that 
       $\lla x,y\rra$ is topologically closed $\forall x,y\in \Gamma$.}

\noindent
  Then $\Gamma$ is isomorphic to the interval $[0,1]\subseteq\Reals$ by a
  simultaneous order- and topological isomorphism.
\smallskip}

{\narrower\smallskip\noindent
{\bf Proof:}
Let $S$ be the countable (topologically) dense subset.  To establish that
$\Gamma$ is order-isomorphic  to
$[0,1]\subseteq\Reals$, we verify the hypotheses of Theorem 5 by first
checking that $S$ is order-dense and then verifying condition (iii) of
that theorem.

First, to prove that $S$ is order-dense, observe that any non-empty
order-open interval $I =\la x,y \ra$ is also topologically open, because
its complement, $\lla 0, x\rra \cup \lla y, 1\rra$, is the union of two
order-closed intervals, and order-closed intervals are also topologically
closed by hypothesis.  Thus $I$ contains points of $S$ as required.

To verify condition (iii) of Theorem 5, let us first show that any ``past
set'' (or ``order ideal'') in $\Gamma$ has a supremum (a past set being a
subset $A$ such that $x\prec y \in A \implies x\in A$).  To that end,
observe first that, if $A$ is a past set, then $A$, as ordered by $\prec$,
is a directed set and therefore defines a {\it net} in the topological
space $\Gamma$.  As $\Gamma$ is compact this net has an accumulation point
$a$, which we will show is the desired supremum.  In fact, if $x\in A$ then
by definition all of the points of the net $A$ eventually $\succ x$, {\it
i.e.}  they belong to $\lla x,1\rra$; hence their accumulation point $a$
belongs to $\lla x,1\rra$, because the latter set is topologically closed
by hypothesis, and a closed set contains all of its accumulation points.
Thus $\forall x\in A$, $x\prec a $, and we conclude that $a$ is an upper
bound of $A$.  On the other hand if $c$ is an arbitrary upper bound of $A$,
then $A\subseteq \lla 0,c \rra$ which, being closed, must contain the
accumulation point $a$ as before, whence $a\prec c$.  Hence, $a$ is the
least upper bound of $A$, as promised.

Now let $\Gamma = A \sqcup B$ with $A \prec B$.  Since $A$ is manifestly a
past set in this situation, it has a supremum $a$, as we have just seen;
and dual reasoning tells us that $B$ also has an infimum $b$.  Then either
$a=b$ or not.

In the former case, either $a=b\in A$ and the first alternative of
condition (iii) holds, or $a=b\in B$ and the second alternative of
condition (iii) holds.

On the other hand if $a$ and $b$ are distinct, then a jump occurs: $A=\lla
0,a \rra$ and $B = \lla b,1 \rra$, which means that $\Gamma$ would be the
disjoint union of a pair of order-closed intervals.  But, as order-closed
intervals are topologically closed by hypothesis, this would mean that
$\Gamma$ was topologically disconnected, contrary to assumption.

So far we have shown that $\Gamma$ is order-isomorphic to
$[0,1]\subseteq\Reals$.  Now we must prove that it is not only tonomorphic
to $[0,1]$, but homeomorphic as well.  To that end, fix a tonomorphism
$f:\Gamma\to[0,1]$ and let $[r,s]\subseteq[0,1]\subseteq\Reals$ be a closed
real interval.  We have $f^{-1}([r,s])=\lla f^{-1}(r),f^{-1}(s)\rra$, which
is closed by hypothesis.  Now the family of all closed intervals in $[0,1]$
generates the topology of $[0,1]$ and, since the inverse images of these
intervals under $f$ are all closed in $\Gamma$, $f$ is continuous.  But a
continuous bijection between compact spaces is a homeomorphism, so $f$ is a
joint homeo- and tonomorphism of $\Gamma$ to $[0,1]\subseteq\Reals$, as
required.  $\quad\dal$\smallskip}

\vskip 2 true cm
\cent{\bf Appendix B: Comparison with Definitions Used in the $C^2$ Context}
\noindent
In a $C^2$-lorentzian manifold every point possesses a so-called convex
normal neighborhood, and the framework presented in references such as
[3], [7] and [8] becomes applicable.  Since $C^2$ is a
special case of $C^0$, the framework of this paper also applies, 
of course, and we may
compare it with the $C^2$ one.  In fact, although some of the
definitions do differ,  most of the important ones coincide, including the
definitions of causal curve and of globally hyperbolic manifold.  Here we
summarize (in a few cases without proof) the most important agreements and
disagreements we know of.  Unless otherwise stated, any terms which we use
without definition will have the meaning assigned to them in ref. [7];
and {\it the manifold in question will always be taken to be
$C^2$-lorentzian as a standing assumption}.

The most important disagreements in definition stem from our use of $K^+$
where the corresponding $C^2$ definition would use $J^+$, the latter being
in general a proper subset of the former.  However in many situations
(including in all globally hyperbolic spacetimes) 
these two causal relations coincide.  In the $C^2$ context, $J^+(p)$ denotes
the set of points reached from $p$ by smooth curves with future-directed
null or timelike tangent; and $\M$ is said to be {\it causally simple}
iff $J^+({\cal K})$ and $J^-({\cal K})$ are closed for every compact set
${\cal K}\subseteq\M$.  (Equally, $\M$ is causally simple iff $J^+(p)$ is
the closure of $I^+(p)$ and $J^-(p)$ is the closure of $I^-(p)$ for each
$p$.)  When $\M$ is causally simple, there is no difference between $J^+$
and $K^+$:

{\narrower\smallskip\noindent
{\bf Lemma 25:}
 In a causally simple $C^2$-lorentzian manifold, 
 $K^+(p)=J^+(p)$ and
 $K^-(p)=J^-(p)$.
\smallskip} 

{\narrower\smallskip\noindent
{\bf Proof:}
Write $p\le q \Leftrightarrow q\in J^+(p)$.  Trivially, $\le$ is transitive
and contains the chronology relation $I^+$ so we check closure.  Let
$x_n\le y_n$ for each $n$, where $x_n\to x$ and $y_n\to y$.  Let us prove
by contradiction that $x \le y$.  If not, then $x \notin J^-(y)$, whence,
since $J^-(y)$ is closed by causal simplicity, $x$ would possess a compact
neighborhood ${\cal K}$ disjoint from $J^-(y)$, whence $y\notin J^+({\cal
K})$.  But since $J^+({\cal K})$ is closed, $y$ would have to have an open
neighborhood disjoint from it, contradicting the existence of timelike
curves from $x_n$ to $y_n$ for arbitrarily large $n$.  Hence $x\le y$ and
$\le$ is closed.  Lastly, since by causal simplicity $J^+(p)$ is the
closure of $I^+(p)$, we cannot find a smaller closed
 relation than $\le$ which
includes $I^+$.  Therefore, $\le$ is $\prec$, and $K^+(p)=J^+(p)\ \forall
p$.  Likewise, $J^-(p)=K^-(p)$.  
$\quad\dal$\smallskip}

Next consider the definition of causal curve.  Not everyone phrases this in
precisely the same manner, but it is easy to verify that, for example, the
definition given in [7] coincides with our Definition 17, as one sees by
noting that $\gamma(t) \prec_{\cal O} \gamma(t')$ in  Definition 17 is 
equivalent to $\gamma(t')\in J^+(\gamma(t),\O)$ by the causal simplicity of 
convex normal neighborhoods.

Now consider the definition of global hyperbolicity, a concept made by
joining together a causality condition with a compactness requirement.
Within our framework, global hyperbolicity is defined in terms of $K^+$ and
signifies that cycles are absent and intervals are compact (Definition 11).
Within the $C^2$ framework, the accepted definition (in one of several
equivalent forms which are in use) refers to $J^+$ rather than $K^+$.  With
that difference, it again requires that intervals be compact, but the
acyclicity condition (intrinsically weaker in the case of $J^+$) is
strengthened to local causal convexity.  (Usually one says ``strong
causality'', but that condition is nothing but local causal convexity with
respect to $J^+$ (or equivalently $I^+$).)  In other words, the $C^2$
framework uses the

{\narrower\smallskip\noindent
{\bf Alternative Definition 26:}
   The $C^2$-lorentzian manifold $\M$ is globally hyperbolic iff it is
   locally $J$-convex and $J(p,q)$ is compact $\forall \, p, q \in \M$.
\smallskip}

\noindent
In defining global hyperbolicity of an open {\it subset} $\O\subseteq\M$,
one requires as well that $\O$ be $J$-convex.

To compare this with Definition 11, we must in particular compare
$K$-causality with 
local $J$-convexity.  But since $I^+\subseteq K^+$ we have as a simple
corollary of Lemma 16:

{\narrower\smallskip\noindent
{\bf Lemma 27:} 
           If ${\cal M}$ is $K$-causal, then it is locally $J$-convex.
\smallskip}
{\narrower\smallskip\noindent
{\bf Proof:}
Apply Lemma 16 and observe that causal convexity relative to $K^+$
immediately implies causal convexity relative to $I^+$.
$\quad\dal$\smallskip}

\noindent
The converse of Lemma 27 is not true.  For example the manifold depicted in
Figure 2c is locally $J$-convex, but it is not $K$-causal because $p\prec q$
and $q\prec p$.  Thus $K$-causality is strictly stronger than ``strong
causality.''\footnote{*}
{R. Low [11] has pointed out to us that the
causal relation known as ``Seifert's $J^+_S$'' is closed and
transitive (and contains $I^+$).  Therefore it contains $\prec$.
Since the condition that $J^+_S$ be a partial order is known as
``stable causality,'' it follows that stable causality implies
$K$-causality.  Whether these two conditions are actually equivalent
is an interesting question.  Both $K$-causality and stable causality 
fail on the manifold of Figure 2c.}

Using Lemma 27, we can prove that global hyperbolicity of $\M$ in our
sense coincides with the usual meaning (which we will refer to as being
globally hyperbolic ``in the $C^2$ sense'').

{\narrower\smallskip\noindent
{\bf Lemma 28:}
     A $C^2$-lorentzian manifold ${\cal M}$ is globally hyperbolic in the
    sense of Definition 26 iff it is globally hyperbolic in the
    sense of Definition 11. 
\smallskip}

{\narrower\smallskip\noindent
{\bf Proof:}
First assume that $\M$ is globally hyperbolic in the $C^2$ sense, in which
case it is also causally simple.  Then, by Lemma 25, $J^\pm=K^\pm$.  Hence
$K(p,q) = J(p,q) := J^+(p)\cap J^-(q)$, which is compact by Definition 26.
Furthermore, $\M$ is $K$-causal
since if there were unequal points which $K$-preceded each other, then they
would also $J$-precede each other (since $J=K$), contradicting the 
local $J$-convexity
 of $\M$.  Hence $\M$ is globally hyperbolic in our sense.

Conversely assume that $\M$ fulfills the conditions of Definition 11. Then
it is certainly locally $J$-convex
by Lemma 27, so we only have to prove that
$J(p,q)$ is compact for any points $p,q\in\M$.
Since $J(p,q)$ is included in the compact set $K(p,q)$, it suffices to show
that it is closed.  Then  let $x_j\to x$ be any convergent sequence with
$x_j\in J(p,q)$ for all $j$.  For each $j$ there is by definition a causal
path $\gamma_j$ such that 
 $\gamma_j(0)=p$ and
$\gamma_j(1)\to x$.  Theorem 23 and Lemma 22 then guarantee us a causal
curve $\Gamma$ from $p$ to $x$, proving that $x\in J^+(p)$, as required.
(Recall that our definition of causal curve coincides with the $C^2$
definition.) 
  Similarly, $x\in J^-(q)$, and we are done.  
$\quad\dal$\smallskip}

\noindent
Because of this equivalence, we have not, in the main text, introduced any
new word to differentiate our concept of global hyperbolicity from the
$C^2$ one.  We should point out, however, that the equivalence demonstrated
in Lemma 28 applies to global hyperbolicity of an entire manifold $\M$, and
not necessarily to arbitrary open subsets of $\M$, the reason being that
for a subset of $\O\subseteq\M$, $J$-convexity does not always imply
$K$-convexity.  (Once again Figure 2c provides a counterexample.)

Finally, consider the definition of convergence for sequences of causal
curves.  In the $C^2$-lorentzian context, at least two distinct notions of
convergence are in use, as summarized in [3], and both can be phrased
in topological terms.  The first is convergence with respect to what
reference [5] calls the ``upper topology,'' and the second is
convergence with respect to the ``lower topology.''    In the upper
topology (sometimes called the $C^0$ 
topology), a sequence (or net) of curves $\Gamma_n$ converges to 
the curve $\Gamma$ iff the $\Gamma_n$ are eventually {\it included} in
every open set 
$\O\subseteq \M$ which includes $\Gamma$ itself.  In the lower topology,
the same sequence (or net) converges to $\Gamma$ iff the $\Gamma_n$
eventually {\it meet} (have non-empty intersection with) every open set
$\O\subseteq \M$ which meets $\Gamma$.

The upper and lower topologies are essentially equivalent in locally
$J$-convex manifolds (Prop. 2.21 in [3]), but taken separately they
have 
the drawback of not being Hausdorff unless supplemented by conditions
controlling the endpoints of the curves (see for example [8]).  By
requiring both types of convergence at once, we obtain a topology which is
automatically Hausdorff, and is in fact the Vietoris topology, or more
precisely the topology on the space of causal curves induced by the
Vietoris topology on $2^\M$.  In this paper we have interpreted convergence
in the sense of Vietoris because, by doing so, we were able to prove the
existence of accumulation curves by first constructing accumulation {\it
sets}, and then showing that these sets are in fact causal curves.

%
\vbox{%
\includegraphics{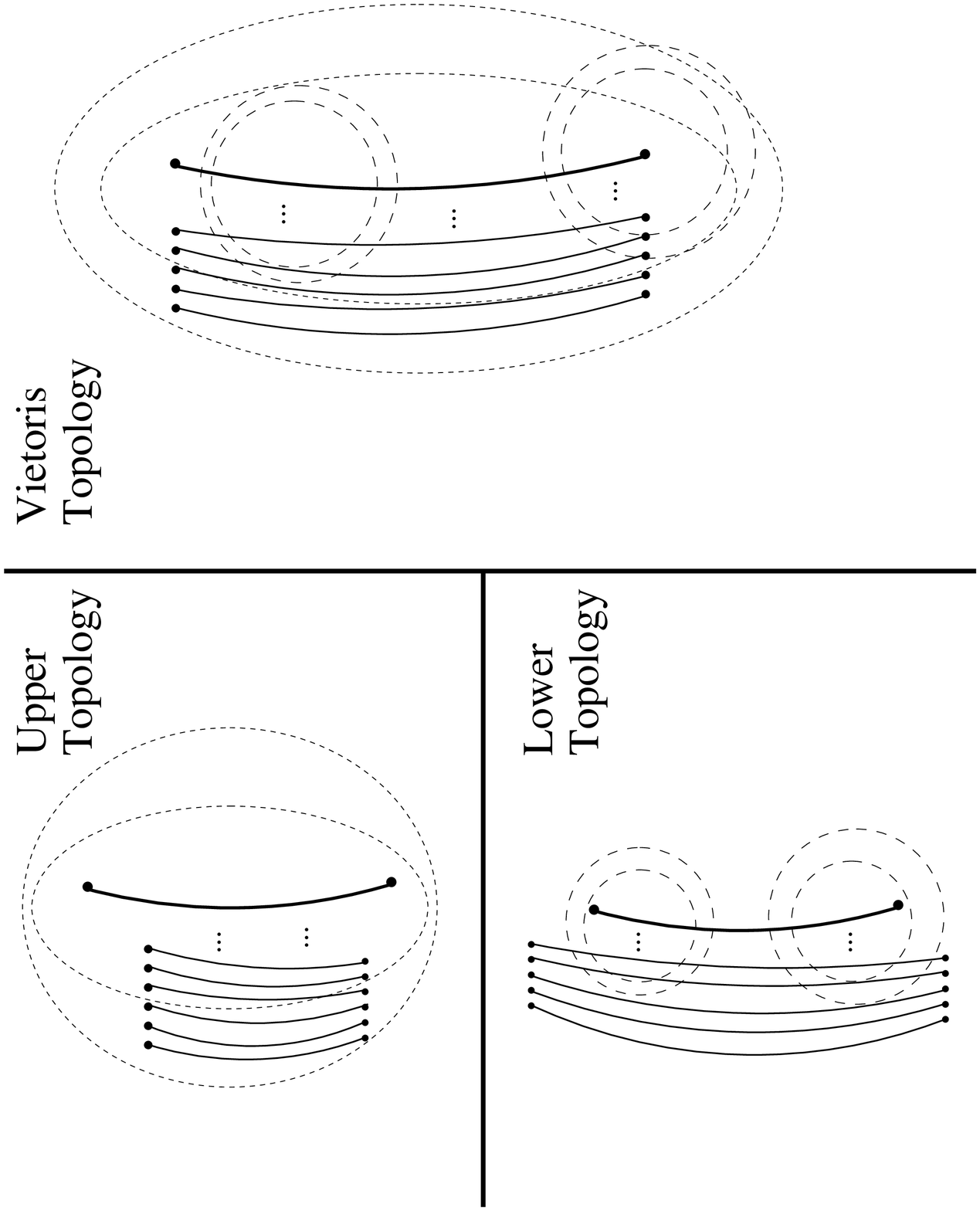}
\vskip 110 true mm
{\narrower\narrower\singlespace\eightrm\bigskip
\textfont0=\eightrm
\textfont1=\eighti \scriptfont1=\sixi
\textfont2=\eightsy \scriptfont2=\sixsy
\noindent
{\underbar {Figure~3:}}
On the upper-left, the sequence of curves depicted converges in the
upper topology to the curve represented by the heavy line, since every 
${\cal M}$-open
set containing the latter also contains all but finitely many curves of the
sequence.  The endpoints of the sequence curves need not converge to the
endpoints of the limit curve.

The sequence of curves depicted on the lower-left converges in the
lower topology to the curve represented by the heavy line, since 
every ${\cal M}$-open set that meets the latter also meets all but finitely
many curves of the sequence.  Again, without endpoint conditions, the
endpoints of the sequence curves need not converge to the endpoints
of the limit.

The right-hand sequence of curves converges in the Vietoris topology to the
curve depicted by the heavy line, since the sequence meets both the
criterion for convergence in the upper topology and the criterion for
convergence in the lower topology.  Notice that convergence of the
endpoints is enforced automatically (Lemma 22).
\bigskip\bigskip}}
%

Now in general the upper, lower, and Vietoris topologies all differ, but
in a locally $J$-convex region, they induce precisely the same topology on
the space $C(p,q)$ of causal curves from a given point $p$ to a given
point $q$; and this is all that is relevant in comparing our Theorem 23
with the corresponding theorem from the $C^2$ context.  In fact, all we
need in order to show that the $C^2$ theorem follows from ours is the
trivial observation that the Vietoris topology is in every situation
either equal to or finer than the upper and lower topologies.

Having taken note of this, 
we can easily derive the standard $C^2$ theorem as a corollary of
Theorem 23.  In other words, we can demonstrate that within any open subset
$\O\subseteq\M$ which is globally hyperbolic in the standard $C^2$ sense
(as described after 
Definition 26), the space of curves $C(p,q)$ is compact
with respect to the upper topology.  To see this, notice first that $\O$,
regarded as a manifold in its own right, will be globally hyperbolic in the
sense of Definition 26, and therefore also globally hyperbolic in our
sense, by Lemma 28.  But then Theorem 23 implies that $C(p,q)$ is compact
in the Vietoris topology, hence compact in any weaker topology, including
in particular the upper topology. 

\par\vfil\eject\bye